\documentclass[12pt]{article}

\renewcommand{\thefootnote}{\fnsymbol{footnote}}

\usepackage{amsbsy,amssymb,latexsym,amsfonts,amsmath}
\usepackage{mathrsfs}
\usepackage{cite}
\usepackage{bm}
\usepackage{here}
\usepackage{comment}
\usepackage{amsmath}
\usepackage{cases}
\usepackage {empheq}
\usepackage{color}
\usepackage[pdftex]{graphicx}
\usepackage{braket}
\usepackage{slashed}
\usepackage{cancel}
\numberwithin{equation}{section}
\makeatletter
\def\EqNumText{\refstepcounter{equation}\cdots\tagform@\theequation}%
\makeatother

\newcommand{\bel}[1]{\begin{equation}\label{#1}}                     
\newcommand{\bal}[1]{\begin{eqnarray}\label{#1}}                     
\newcommand{\be}{\begin{equation}}
\newcommand{\ee}{\end{equation}}

\begin{document}
%
%
\begin{titlepage}
\begin{flushright}
\normalsize
~~~~
NITEP 121\\
OCU-PHYS 549\\
October 19, 2021\\
\end{flushright}

\vspace{15pt}

\begin{center}
{\Large  Target Space Duality of Non-Supersymmetric String Theory} \\
\end{center}

\vspace{23pt}

\begin{center}
	{ H. Itoyama$^{a, b,c}$\footnote{e-mail: itoyama@osaka-cu.ac.jp}, Yuichi Koga$^b$\footnote{e-mail: k-yuichi@yj.osaka-cu.ac.jp} and Sota Nakajima$^b$\footnote{e-mail: nakajima@rx.osaka-cu.ac.jp}   }\\
	
	%
	\vspace{10pt}
	%
	
	$^a$\it Nambu Yoichiro Institute of Theoretical and Experimental Physics (NITEP),\\
	Osaka City University\\
	\vspace{5pt}
	
	$^b$\it Department of Mathematics and Physics, Graduate School of Science,\\
	Osaka City University\\
	\vspace{5pt}
	
	$^c$\it Osaka City University Advanced Mathematical Institute (OCAMI)
	
	\vspace{5pt}
	
	3-3-138, Sugimoto, Sumiyoshi-ku, Osaka, 558-8585, Japan \\

\end{center}
%
\vspace{15pt}
\begin{center}
Abstract\\
\end{center}
T-dualities of the non-supersymmetric string models, which are constructed by twisted compactifications, are investigated. We show that the T-duality groups of such models are obtained by imposing congruence conditions on $O\left( d_{L},d_{R},\mathbb{Z}\right) $, that is, the non-supersymmetric string models are invariant under congruence subgroups of $O\left( d_{L},d_{R},\mathbb{Z}\right)$. We also point out that the transitions among the non-supersymmetric models can be induced by acting $O\left( d_{L},d_{R},\mathbb{Z}\right) $ transformations.
 

\vfill

\end{titlepage}

\renewcommand{\thefootnote}{\arabic{footnote}}
\setcounter{footnote}{0}


\section{Introduction and conclusions}	
It is known that string theory often gives us new properties of target space that are not found in theory of point particles. One of the simplest examples is the circle compactification of bosonic string theory; bosonic strings on a circle with a radius $R$ are equivalent to those on a circle with $\alpha'/R$  \cite{Kikkawa:1984cp,Sakai:1985cs}. If compact target space is higher-dimensional, we can see a variety of such discrete symmetries which are called target space dualities or T-dualities. In general, strings on a $d$-dimensional torus have the $O(d_{L},d_{R},\mathbb{Z})$ duality symmetry where $d_{L}=d_{R}=d$ for bosonic or type II string theory and $d_{L}-16=d_{R}=d$ for heterotic string theory \cite{Giveon:1988tt,Giveon:1994fu}. 
T-duality is a unique nature in string theory and yields a lot of interesting possibilities.
For instance, some elements of T-dualities allow us to consider non-geometric backgrounds such as asymmetric orbifolds or T-folds \cite{Narain:1986qm,Hellerman:2002ax,Dabholkar:2002sy,Flournoy:2004vn,Hull:2004in,Hull:2006va}. One of the benefits of considering strings on non-geometric backgrounds is that the number of unfixed moduli is fewer than on geometric backgrounds. Hence, the string models with non-geometric backgrounds are often adopted for realistic model building or clarifying interesting features of strings (see e.g. \cite{Ibanez:1987pj,Taylor:1987uv,Erler:1996zs,Aoki:2004sm,Tan:2015nja,Satoh:2015nlc,Sugawara:2016lpa,GrootNibbelink:2017usl,Aoyama:2021kqa}). Furthermore, an attempt to construct a field theory that incorporates T-duality symmetries including non-geometric ones as manifest symmetries has been proposed and studied \cite{Hull:2009mi,Hohm:2010jy,Aldazabal:2013sca}. 
In the context of a bottom-up approach, non-Abelian discrete symmetries (e.g. the modular symmetry), which can be regarded as (parts of) T-dualities, are considered as a candidate for an origin of the flavor symmetry in the standard model \cite{Kobayashi:2006wq,Ishimori:2010au,Altarelli:2010gt,deAdelhartToorop:2011re,Hernandez:2012ra,King:2013eh,King:2014nza,Kikuchi:2020nxn,Ishiguro:2020tmo}. 

Toroidal compactifications of superstring theory, which provide the $O(d_{L},d_{R},\mathbb{Z})$ duality symmetry, preserve the maximal supersymmetries. However, the supersymmetry has not been found at least around the current achievable energy scale by accelerator experiments. Considering this fact, it is worth devoting attention to the top-down scenario that supersymmetry is already broken at a very high energy scale like the Planck/string scale. In this paper, we focus on the string models in which the supersymmetry is broken by a stringy version of the Scherk-Schwarz mechanism (twisted compactification) \cite{Spontaneous Breaking of Supersymmetry Through Dimensional Reduction,Spontaneous Supersymmetry Breaking in Supersymmetric String Theories,Kounnas:1989dk}. The construction of such models in ten-dimensions was originally proposed in \cite{Dixon:1986iz,AlvarezGaume:1986jb}, and extended to the general dimensional cases in \cite{Nair:1986zn,Ginsparg:1986wr}. One of the great advantages of such non-supersymmetric models is to allow the exponential suppression of the cosmological constant in a particular region of the moduli space \cite{Itoyama:1986ei,Itoyama:1987rc}. For avoiding the problem of the vacuum instability, it is favorable to set the non-supersymmetric models whose cosmological constants are zero or very small as starting points, and therefore such string models are often considered in a top-down approach in which the supersymmetry is supposed to be broken at a very high energy scale \cite{Abel:2015oxa,Aaronson:2016kjm,Abel:2017rch,Blum:1997gw,Kounnas:2015yrc,Kounnas:2016gmz,Kounnas:2017mad,Coudarchet:2017pie,Coudarchet:2018ztz,Partouche:2018ftj,Florakis:2016ani,Abel:2017vos,Itoyama:2019yst,Itoyama:2020ifw,Itoyama:2021fwc,Partouche:2019pgv,Angelantonj:2019gvi,Abel:2020ldo,Coudarchet:2020sjw,Coudarchet:2021qwc,Faraggi:2009xy,Faraggi:2019fap,Faraggi:2020fwg,Faraggi:2020hpy,Faraggi:2020wej,Faraggi:2020wld,Faraggi:2021mws,Ashfaque:2015vta,Itoyama:2021kxp}. 

The purpose of this work is to identify the T-duality groups of the non-supersymmetric models constructed by the twisted compactifications. The outline of this paper is as follows. We devote ourselves to review the construction of the non-supersymmetric string models in section \ref{non-susy string}. There are a variety of different non-supersymmetric models, depending on the choices of Narain lattice twists of order 2. We also give the one-loop partition functions of the non-supersymmetric models, which help us to specify the T-dualities and the free-spectra. In section \ref{T-duality of non-susy}, we identify the T-duality groups of the non-supersymmetric models by considering those of the toroidal models and the construction of the non-supersymmetric models which is introduced in the previous section. 
In subsection \ref{T-duality type II}, we give the example of the T-duality symmetry in the type II case. In particular, we focus on the T-duality groups with $d=2$, in which the basis of the moduli space can be chosen to have two modular symmetries.
In subsection \ref{T-duality heterotic}, we turn to the T-duality groups in the heterotic case. At the end of this paper, we discuss the gauge symmetry enhancement in the non-supersymmetric heterotic models, in connection with the T-duality symmetries.

We here briefly summarize the main results of this work. Not all elements of $O(d_{L},d_{R},\mathbb{Z})$ which is the T-duality group of the toroidal models are symmetries in the non-supersymmetric models. In order for the elements of $O(d_{L},d_{R},\mathbb{Z})$ to be symmetries after the supersymmetry is broken, they have to satisfy a congruence condition that depends on the choice of a shift-vector. As a result, the T-duality groups of the non-supersymmetric models form congruence subgroups of $O(d_{L},d_{R},\mathbb{Z})$. Moreover, we see that acting an element of $O(d_{L},d_{R},\mathbb{Z})$ on the non-supersymmetric model whose duality group does not include the element makes a transition to another non-supersymmetric model. As concrete examples, we study the T-duality groups in the type II case with $d=2$ and the heterotic case with $d=1$. We obtain the similar results that have been observed in \cite{Gregori:1997hi}.
In the type II case with $d=2$ in which the T-duality group $O(2,2,\mathbb{Z})$ of the toroidal models can be decomposed into $PSL(2,\mathbb{Z})\times PSL(2,\mathbb{Z})$, the Hecke congruence subgroups and the theta subgroup of the modular group are found in the T-duality groups of the non-supersymmetric models. In the heterotic case with $d=1$, we classify the possible non-supersymmetric models into four classes and identify which elements of $O(17,1,\mathbb{Z})$ can be symmetries for each of the four classes. In particular, we see that the restrictions of parameters of the (dual) Wilson line shifts are different depending on the classes that the non-supersymmertic models belong to. Moreover, we naively suggest the relations between the gauge symmetry enhancements and the T-duality symmetries.

\section{Non-supersymmetric strings}\label{non-susy string}

\subsection{Construction}\label{construction}
In this section, we review how to construct non-supersymmetric string models by twisted compactifications (for details see \cite{Dixon:1986iz,Ginsparg:1986wr}). The starting point is a toroidal model $d$-dimensional compactified in which supersymmetry is maximally preserved. The pairings of states in the toroidal models are 
\begin{align}
\text{Type IIB string:}&~\left( \Gamma^{d,d}; v \bar{v},s\bar{s},v\bar{s},s\bar{v}\right), \\
\text{Heterotic string:}&~\left( \Gamma^{16+d,d};\bar{v},\bar{s}\right),
\end{align}
where $\Gamma^{d_{L},d_{R}}$, called a Narain lattice, is an even self-dual lattice with Lorentzian signature $(d_{L},d_{R})$ and $o$, $v$, $s$ and $c$ represent the conjugacy classes of $SO(8)$ (see appendix \ref{appendixA}). The non-supersymmetric model is constructed by orbifolding the toroidal model by a $\mathbb{Z}_2$ shift action that gives an eigenvalue $(-1)^{F}e^{2\pi i\delta\cdot p}$ for a state with an internal momentum $p$. Here $F$ is the spacetime fermion number and $\delta$ is a shift-vector in the Narain lattice such that $2\delta\in \Gamma^{d_{L},d_{R}}$.
It is convenient to split $\Gamma^{d_{L},d_{R}}$ into two subsets $\Gamma^{d_{L},d_{R}}_{+}$ and $\Gamma^{d_{L},d_{R}}_{-}$ as follows:
\begin{align}\label{Gammapm}
\Gamma_{+}^{d_{L},d_{R}}(\delta)=\left\lbrace \left.  p\in \Gamma^{d_{L},d_{R}}~\right|  \delta\cdot p\in \mathbb{Z} \right\rbrace,~~~~~
\Gamma_{-}^{d_{L},d_{R}}(\delta)=\left\lbrace \left.  p\in \Gamma^{d_{L},d_{R}} ~\right| \delta\cdot p \in \mathbb{Z} +\frac{1}{2}\right\rbrace.
\end{align}
We shall often denote $\Gamma_{\pm}^{d_{L},d_{R}}$ omitting the argument $\delta$ for simplicity.
Then, in the untwisted sectors, after modding out by the $\mathbb{Z}_{2}$ shift action, the surviving pairings of states are 
\begin{align}
&\text{Type IIB string:}~\left( \Gamma^{d,d}_{+};v\bar{v}\right), \left( \Gamma^{d,d}_{+};s\bar{s} \right), \left( \Gamma^{d,d}_{-};v\bar{s}\right), \left( \Gamma^{d,d}_{-};s\bar{v}\right),\\
&\text{Heterotic string:}~\left( \Gamma^{16+d,d}_{+};\bar{v}\right),\left( \Gamma^{16+d,d}_{-};\bar{s}\right).
\end{align}
Modular invariance of the one-loop partition function requires that $\delta^{2}$ be an integer and the twisted sectors be added. For $\delta^2$ odd, the pairings of states in the twisted sectors are
\begin{align}
&\text{Type IIB string:}~\left( \Gamma^{d,d}_{-}+\delta;o\bar{o},c\bar{c}\right),  \left( \Gamma^{d,d}_{+}+\delta;o\bar{c} ,c\bar{o}\right),\\
&\text{Heterotic string:}~\left( \Gamma^{16+d,d}_{+}+\delta;\bar{o}\right),\left( \Gamma^{16+d,d}_{-}+\delta;\bar{c}\right),
\end{align}
and for $\delta^2$ even,
\begin{align}
&\text{Type IIB string:}~\left( \Gamma^{d,d}_{+}+\delta;o\bar{o},c\bar{c}\right), \left( \Gamma^{d,d}_{-}+\delta;o\bar{c}, c\bar{o}\right),\\
&\text{Heterotic string:}~\left( \Gamma^{16+d,d}_{-}+\delta;\bar{o}\right),\left( \Gamma^{16+d,d}_{+}+\delta;\bar{c}\right).
\end{align}
With $d=0$ we obtain the 10D non-supersymmetric string models, i.e. the type 0B model in the type IIB case, or the non-supersymmetric heterotic models originally constructed in \cite{Dixon:1986iz,AlvarezGaume:1986jb,Ginsparg:1986wr} in the heterotic case.
The non-supersymmetric models with the toroidal type IIA model being a starting point are obtained by flipping the chirality of the right-moving spinors of $SO(8)$ in the type IIB case.

\subsection{Partition function}\label{partition function}

Possible Narain lattices are characterized by a set of parameters $\lambda^{a}$ called moduli. We can introduce the generalized vierbein $\tilde{\mathcal{E}}(\lambda^{a})$ of the Narain lattice, which is expressed as a $(d_L+d_R)\times (d_L+d_R)$ matrix. In order for the Narain lattice to be even and self-dual, the Narain metric which is defined as $J=\tilde{\mathcal{E}}\eta\tilde{\mathcal{E}}^t$ with $\eta=diag\left(\boldsymbol{1}_{d_{L}},-\boldsymbol{1}_{d_{R}} \right)$ must be an integer matrix with signature $\left(d_{L},d_{R} \right)$ of which diagonal components are even and determinant is $\pm1$. 
Then, an element $p$ of the Narain lattice is written as 
\begin{align}
p=Z\tilde{\mathcal{E}}(\lambda^{a}),
\end{align}
where $Z$ is a $(d_{L}+d_{R})$-dimensional row vector with integer components. Note that the inner product of $p_{1}=Z_{1}\tilde{\mathcal{E}}$ and $p_{2}=Z_{2}\tilde{\mathcal{E}}$ is independent of the moduli $\lambda^{a}$:
\begin{align}
p_{1}\cdot p_{2}=Z_{1}\tilde{\mathcal{E}}(\lambda^{a})\eta \tilde{\mathcal{E}}^{t}(\lambda^{a})Z_{2}^{t}=Z_{1}JZ_{2}^{t}.
\end{align}

The one-loop partition function of the toroidal string models can be written as
\begin{align}\label{toroidal model}
Z^{T^d}(\lambda^a)=Z_{B}^{(8-d)}Z_{F}Z_{\Gamma^{d_L,d_R}}(\lambda^a).
\end{align}
Here the individual contributions to the partition function are\footnote{We omit the modular parameter $\tau$ of the world-sheet torus from the arguments of the partition function.} 
\begin{align}
&Z_{B}^{(8-d)}=\tau_{2}^{\frac{8-d}{2}}\left(\eta\bar{\eta} \right)^{-(8-d)},\\
&Z_{F}=\begin{cases}
\left( V_{8}-S_{8}\right) \left( \bar{V}_{8}-\bar{S}_{8} \right)~\text{or}~\left( V_{8}-S_{8}\right) \left( \bar{V}_{8}-\bar{C}_{8} \right)&\text{(type IIB or type IIA string)}\\
\bar{V}_{8}-\bar{S}_{8}&\text{(heterotic string)}
\end{cases},\\
&Z_{\Gamma^{d_L,d_R}}=\eta^{-d_{L}}\bar{\eta}^{-d_{R}}\sum_{p\in \Gamma^{d_{L},d_{R}}}q^{\frac{1}{2}p_{L}^{2}}\bar{q}^{\frac{1}{2}p_{R}^{2}},~~~\text{with}~
\begin{cases}
d_{L}=d_{R}=d&\text{(type II string)}\\
d_{L}-16=d_{R}=d&\text{(heterotic string)}
\end{cases},
\end{align}
where $q=e^{2\pi i \tau}$, $\eta(\tau)$ is the Dedekind eta function and $\left(O_{8}, V_{8},S_{8},C_{8}\right)$ denotes a set of the characters of $SO(8)$ (see appendix \ref{appendixA}). Note that the partition function is invariant under the rotations $O(d_{L},\mathbb{R})\times O(d_{R},\mathbb{R})$ which act on the left- and right-moving momenta respectively. 

As mentioned above, the non-supersymmetric model is constructed by splitting the Narain lattice $\Gamma^{d_{L},d_{R}}$ by a shift-vector $\delta$ accompanied with $(-1)^{F}$. Since $2\delta$ is in $\Gamma^{d_{L},d_{R}}$, the shift-vector $\delta$ is expressed as
\begin{align}
\delta=\frac{1}{2}\hat{Z}\tilde{\mathcal{E}}(\lambda^{a}),
\end{align}
for a certain integer vector $\hat{Z}\in\mathbb{Z}^{d_{L}}\oplus \mathbb{Z}^{d_{R}}$. Recalling that $\delta^{2}$ is required to be an integer, $\hat{Z}$ must satisfy
\begin{align}\label{delta2integer}
\hat{Z}J\hat{Z}^t=0~(\text{mod 4}).
\end{align}
The non-supersymmetric models are characterized by a set of integers $\hat{Z}$ that satisfies \eqref{delta2integer}. As we shall discuss later, there are conditions other than \eqref{delta2integer} imposed on inequivalent choices of $\hat{Z}$. It is convenient to denote the shift-vector and the partition function as $\delta_{(\hat{Z})}$ and $Z^{\cancel{SUSY}}_{(\hat{Z})}$, for the choice of $\hat{Z}$ to be clear.
Let us write down the partition function $Z^{\cancel{SUSY}}_{(\hat{Z})}$. 
Following the construction in subsection \ref{construction}, the partition function in the type IIB case is
\begin{align}\label{non-susy typeiiB}
Z^{\cancel{SUSY}}_{(\hat{Z})}(\lambda^a)=Z_{B}^{(8-d)}&\left\lbrace \left(V_{8}\bar{V}_{8} +S_{8}\bar{S}_{8}\right) Z_{\Gamma^{d,d}_{+}}(\lambda^a)-\left(V_{8}\bar{S}_{8} +S_{8}\bar{V}_{8}\right) Z_{\Gamma^{d,d}_{-}}(\lambda^a)
\right.\nonumber\\
&\left. +\left(O_{8}\bar{O}_{8} +C_{8}\bar{C}_{8}\right) Z_{\Gamma^{d,d}_{\mp}+\delta} (\lambda^a)-\left(O_{8}\bar{C}_{8} +C_{8}\bar{O}_{8}\right) Z_{ \Gamma^{d,d}_{\pm}+\delta} (\lambda^a)\right\rbrace .
\end{align}
In the heterotic case,
\begin{align}\label{non-susy hetero}
Z^{\cancel{SUSY}}_{(\hat{Z})}(\lambda^a)=Z_{B}^{(8-d)}&\left\lbrace \bar{V}_{8} Z_{\Gamma^{16+d,d}_{+}}(\lambda^a)-\bar{S}_{8} Z_{\Gamma^{16+d,d}_{-}}(\lambda^a)
\right.\nonumber\\
&\left. +\bar{O}_{8} Z_{\Gamma^{16+d,d}_{\pm}+\delta} (\lambda^a)-\bar{C}_{8}  Z_{ \Gamma^{16+d,d}_{\mp}+\delta} (\lambda^a)\right\rbrace .
\end{align}
In the twisted sectors, the upper and lower signs in $Z_{\Gamma^{d_{L},d_{R}}_{\pm}+\delta} $ apply for $\delta^2$ odd and for $\delta^2$ even respectively, which is required for the invariance under $\tau\to\tau+1$ (for details see appendix \ref{check modular inv}).
We see that $Z_{\Gamma^{d_{L},d_{R}}_{\pm}+\delta} $ can be written as
\begin{align}
Z_{\Gamma_{\pm}^{d_L,d_R}+\delta}
=\eta^{-d_{L}}\bar{\eta}^{-d_{R}}\sum_{p\in \Gamma_{\pm}^{d_{L},d_{R}}+\delta}q^{\frac{1}{2}p_{L}^{2}}\bar{q}^{\frac{1}{2}p_{R}^{2}}
=\eta^{-d_{L}}\bar{\eta}^{-d_{R}}\sum_{p\in \Gamma^{d_{L},d_{R}}}\frac{1\pm e^{2\pi i \delta\cdot p}}{2}q^{\frac{1}{2}\left( p_{L}+\delta_{L} \right)^{2}}\bar{q}^{\frac{1}{2}\left( p_{R}+\delta_{R}\right)^{2}}.
\end{align}                                                                                                                                                                                                                                                                                                                                                                                                                                                                                                                                                                                                                                                                                                                                                                                                                                                                                                                                                                                                                                                                                                                                                                                                                                                                                                                                                                                                                                                                                                                                                                                                                                                                                                                                                                                                                                                                                                                                                                                                                                                                                                                                                                                                                                                                                                                                                                                                                                                                                                                                                                                                                                                                                                                                                                                                                                                                                                                                                                                                                                                                                                                                                                                                                                                                                                                                                                                                                                                                                                                                                                                                                                                                                                                                                                                                                                                                                                                                                                                                                                                                                                                                                                                                                                                     

\begin{table}[t]
	\begin{center}
		\begin{tabular}{|c||c|c|c|} \hline
			$\delta=\frac{\hat{\pi}}{2}$ & $\left(1,0^7;0^8 \right)$ & $\left(\left( \frac{1}{2}\right) ^2,0^6;\left( \frac{1}{2}\right) ^2,0^6\right) $ & $\left(1,0^7;1,0^7\right) $ \\\hline
			gauge sym. & $SO(16)\times E_8$ & $\left( E_{7}\times SU(2)\right)^2 $ & $SO(16)\times SO(16)$ \\\hline
		\end{tabular}
		\caption{The 10D non-supersymmetric heterotic models constructed from the $E_{8}\times E_{8}$ lattice and the realized gauge symmetries are listed.}
	\end{center}
\end{table}
\begin{table}[t]
	\begin{center}
		\begin{tabular}{|c||c|c|c|c|} \hline
			$\delta=\frac{\hat{\pi}}{2}$ & $\left(1,0^{15} \right)$ & $\left(\left( \frac{1}{2}\right) ^4,0^{12}\right) $& $\left( \left( \frac{1}{4}\right)^{16}\right) $ & $\left(\left( \frac{1}{2}\right) ^8,0^8\right)$ \\\hline
			gauge sym. & $SO(32)$ & $SO(24)\times SO(8) $ & $SU(16)\times U(1)$ & $SO(16)\times SO(16)$ \\\hline
		\end{tabular}
		\caption{The 10D non-supersymmetric heterotic models constructed from the $Spin(32)/\mathbb{Z}_{2}$ lattice and the realized gauge symmetries are listed.}
	\end{center}
\end{table}                             

We should note that there are equivalent choices of $\hat{Z}$. From the definition  \eqref{Gammapm} of $\Gamma_{\pm}^{d_{L},d_{R}}$, two choices $\hat{Z}_{1}$ and $\hat{Z}_{2}$ give the same splitting of the Narain lattice if $v\in \mathbb{Z}^{d_{L}}\times \mathbb{Z}^{d_{R}}$ exists such that $\hat{Z}_{1}=\hat{Z}_{2}+2v$. Namely, we can only focus on the choices of $\hat{Z}$ in which each of the components takes either 0 or 1, except for $\hat{Z}=\left( 0^{d_{L}+d_{R}}\right) $.
Considering the condition \eqref{delta2integer}, the number of possible choices of $\hat{Z}$ is more restricted. For example, there are three and four 10D non-supersymmetric heterotic models with starting points being the heterotic $E_{8}\times E_{8}$ and $Spin(32)/\mathbb{Z}_{2}$ models respectively, as shown in Table 1 and 2 (for details see \cite{Dixon:1986iz})\footnote{In the $d=0$ case, the shift-vectors are equivalent not only up to shifts by $p_{0}\in\Gamma^{16}$ but also up to permutations of the components since there are no continuous parameters (moduli) to couple to $Z$. }. Note that the shift-vector with $d=0$ is one-half of an element of the $E_{8}\times E_{8}$ or $Spin(32)/\mathbb{Z}_{2}$ root lattice, which is denoted by $\hat{\pi}/2$.

\section{Target space duality of non-supersymmetric strings}\label{T-duality of non-susy}

From now, we assume $d\geq1$ in order to discuss the T-duality groups.
The consistency of string theory requires that the Narain lattice be an even self-dual with Lorentzian signature $\left( d_{L},d_{R}\right)$. Picking up a Narain lattice with a generalized vierbein $\tilde{\mathcal{E}}_{0}$, one can obtain all even self-dual lattices with the same Narain metric by acting $g\in O(d_{L},d_{R},\mathbb{R})$ on $\tilde{\mathcal{E}}_{0}$, where the boost $O(d_{L},d_{R},\mathbb{R})$ is defined in terms of the Narain metric $J$: $gJg^t=J$.  As mentioned in the previous section, the deformations by the individual rotations $O(d_{L},\mathbb{R})\times O(d_{R},\mathbb{R})\subset O(d_{L},d_{R},\mathbb{R})$ do not change the partition function. The moduli space of the toroidal models is therefore locally isomorphic to $O(d_{L},d_{R},\mathbb{R})/O(d_{L},\mathbb{R})\times O(d_{R},\mathbb{R})$ \cite{New Heterotic String Theories in Uncompactified Dimensions $<$ 10,Narain:1986am}.

It is however known that there are discrete subgroups that act on the Narain lattice as an automorphism and keep the toroidal model unchanged. For instance, the partition function \eqref{toroidal model} is invariant under $\tilde{\mathcal{E}}\to g\tilde{\mathcal{E}}$ with $g\in O(d_{L},d_{R},\mathbb{Z})\subset O(d_{L},d_{R},\mathbb{R})$. In other words, for $\lambda^a$ and $g\in O(d_{L},d_{R},\mathbb{Z})$, one can find $\lambda'^a$ such that $\tilde{\mathcal{E}}(\lambda'^a)=g\tilde{\mathcal{E}}(\lambda^a)$ up to the rotations $O(d_{L},\mathbb{R})\times O(d_{R},\mathbb{R})$, and hence the two moduli $\lambda^{a}$ and $\lambda'^{a}$ gives the same toroidal model. Then, the space of inequivalent Narain lattice is $O(d_{L},d_{R},\mathbb{Z})\backslash O(d_{L},d_{R},\mathbb{R})/O(d_{L},\mathbb{R})\times O(d_{R},\mathbb{R})$. The discrete group $O(d_{L},d_{R},\mathbb{Z})$ is called a T-duality group of the toroidal model.

The main goal of this section is to identify T-duality groups of the non-supersymmetric models constructed in the previous section. Namely, we wonder whether $\lambda^a$ and $\lambda'^a$ give equivalent non-supersymmetric models whenever $\lambda'^a$ is related to $\lambda^a$ by the T-duality group $O(d_{L},d_{R},\mathbb{Z})$ of the toroidal model:
\begin{align}\label{question}
Z^{T^d}(\lambda^a)=Z^{T^d}(\lambda'^a)~~ \Longrightarrow ~~Z^{\cancel{SUSY}}_{(\hat{Z})}(\lambda^a)\stackrel{\rm{?}}{=}Z^{\cancel{SUSY}}_{(\hat{Z})}(\lambda'^a).
\end{align}
Recalling that the partition function $Z^{\cancel{SUSY}}_{(\hat{Z})}(\lambda^a)$ is obtained from $Z^{T^d}(\lambda^a)$ by splitting the Narain lattice by $\delta_{(\hat{Z})}(\lambda^{a})$, we easily see that the answer is no. In order for $Z^{\cancel{SUSY}}_{(\hat{Z})}$ to be unchanged, the T-duality group of the non-supersymmetric model must maintain the inner products of any $p\in\Gamma^{d_{L},d_{R}}$ with $\delta_{(\hat{Z})}$ mod 1:
\begin{align}\label{inner product maintained}
\delta\cdot p=\delta\cdot p'~~\text{(mod 1)}~~\text{for any}~p\in \Gamma^{d_{L},d_{R}},
\end{align}
where $p'$ is obtained by acting $g$ on $p$. Inserting $p=Z\tilde{\mathcal{E}}$, $p'=Zg\tilde{\mathcal{E}}$ and $\delta_{(\hat{Z})}=\frac{1}{2}\hat{Z}\tilde{\mathcal{E}}$ into \eqref{inner product maintained} leads us to
\begin{align}\label{congruence condition}
\hat{Z}=\hat{Z}g~~\text{(mod 2)}.
\end{align}
For a choice of $\hat{Z}$, let us define a discrete group $D_{(\hat{Z})}\left(d_{L},d_{R} \right) $ as
\begin{align}\label{non-susy duality}
D_{(\hat{Z})}\left(d_{L},d_{R} \right) =\left\lbrace g\in O(d_{L},d_{R},\mathbb{Z}) \left| ~\hat{Z}g=\hat{Z}~(\text{mod 2})\right. \right\rbrace.
\end{align}
Then $D_{(\hat{Z})}\left(d_{L},d_{R} \right)$ corresponds to the T-duality group of the non-supersymmetric model constructed by using the shift-vector $\delta_{(\hat{Z})}$.
Obviously, $D_{(\hat{Z})}(d_{L},d_{R})$ is a subgroup of $O(d_{L},d_{R},\mathbb{Z})$ since
if $g_{1}$ and $g_{2}$ are elements of $D_{(\hat{Z})}(d_{L},d_{R})$ the product $g_{1}g_{2}$ is also in $D_{(\hat{Z})}(d_{L},d_{R})$:
\begin{align}
\hat{Z}g_{1}g_{2}\stackrel{\rm{mod~2}}{=}\hat{Z}g_{2}\stackrel{\rm{mod~2}}{=}\hat{Z}.
\end{align}
One can furthermore notice that the principal congruence subgroup of level 2 of $O(d_{L},d_{R},\mathbb{Z})$, which is defined as
\begin{align}
\Gamma(2)=\left\lbrace g\in O(d_{L},d_{R},\mathbb{Z}) \left| (g)_{AB}\stackrel{\rm{mod~2}}{=}1~\text{for $A=B$},~~ (g)_{AB}\stackrel{\rm{mod~2}}{=}0~\text{for $A\neq B$}\right. \right\rbrace,
\end{align}
is a subgroup of $D_{(\hat{Z})}(d_{L},d_{R})$. The T-duality group $D_{(\hat{Z})}(d_{L},d_{R})$ is thus a congruence subgroup of $O(d_{L},d_{R},\mathbb{Z})$.

We can understand the above result from a different point of view. Let $\lambda'^{a}$ denote the moduli that are related to $\lambda^{a}$ by $g\in O(d_{L},d_{R},\mathbb{Z})$. The shift-vector $\delta_{(\hat{Z})}(\lambda'^a)$ can be then expressed as a function of $\lambda^{a}$ by using $g$ as follows:
\begin{align}
\delta_{(\hat{Z})}(\lambda'^a)=\frac{1}{2}\hat{Z}\tilde{\mathcal{E}}(\lambda'^a)=\frac{1}{2}\hat{Z}g\tilde{\mathcal{E}}(\lambda^a)u\sim\delta_{(\hat{Z}g)}(\lambda^a),
\end{align}
where $u\in O(d_{L},\mathbb{R})\times O(d_{R},\mathbb{R})$. 
Using $Z^{T^d}(\lambda'^a)=Z^{T^d}(\lambda^a)$ and the invariance of $Z^{T^d}(\lambda^a)$ under $O(d_{L},\mathbb{R})\times O(d_{R},\mathbb{R})$, we get 
\begin{align}\label{duality transition}
Z^{\cancel{SUSY}}_{(\hat{Z})}(\lambda'^a)=Z^{\cancel{SUSY}}_{(\hat{Z}g)}(\lambda^a).
\end{align}
Therefore, in order for the proposition \eqref{question} to be true, it is required that $\hat{Z}g$ be in an equivalent choice to $\hat{Z}$. As mentioned at the end of the previous section, the condition that $\hat{Z}g$ is in an equivalent choice to $\hat{Z}$ is $\hat{Z}g=\hat{Z}$ mod 2.  
Then we get $D_{(\hat{Z})}(d_{L},d_{R})$ defined in \eqref{non-susy duality} as the T-duality group of the non-supersymmetric model.
Eq. \eqref{duality transition} also implies that acting $g$ not in $D_{(\hat{Z})}(d_{L},d_{R})$ on the non-supersymmetric model with the choice $\hat{Z}$ gives another non-supersymmetric model with the choice $\hat{Z}g$. Therefore $g\in O(d_{L},d_{R},\mathbb{Z})$, in general, induces the transitions among the non-supersymmetric models, and the models of which the T-duality groups include $g$ correspond to the fixed points of the transitions.

\subsection{T-duality in the type II case}\label{T-duality type II}

There are $d\times d$ moduli in the toroidal  type II models: a metric $G=ee^t$ of the compactification lattice, an anti-symmetric two-form $B$. Note that these moduli are described by a $d\times d$ matrix $E=G+B$ called a background matrix.
The standard choice of a Narain metric in $\Gamma^{d,d}$ is 
\begin{align}
J=\left(
\begin{array}{cc}
0 & \boldsymbol{1}_{d} \\
\boldsymbol{1}_{d} & 0
\end{array}
\right).
\end{align}
An element of the Narain lattice in the type II case is then given by
\begin{align}
p=Z\tilde{\mathcal{E}}(e,B)=Z\mathcal{E}(e,B)\tilde{\mathcal{E}}_{0},
\end{align}
where 
\begin{align}
\mathcal{E}(e,B)=\left(
\begin{array}{cc}
e & Be^{-t} \\
0 & e^{-t}
\end{array}
\right),~~~
\tilde{\mathcal{E}}_{0}=\frac{1}{\sqrt{2}}\left(
\begin{array}{cc}
\boldsymbol{1}_{d} & -\boldsymbol{1}_{d} \\
\boldsymbol{1}_{d} & \boldsymbol{1}_{d} 
\end{array}
\right).
\end{align}
One can check $\tilde{\mathcal{E}}_{0}\eta \tilde{\mathcal{E}}_{0}^{t}=J$ and $\mathcal{E}J\mathcal{E}^{t}=J$ so that $\tilde{\mathcal{E}}\eta\tilde{\mathcal{E}}^t=J$.
Using the background matrix $E$, $p=\left( p_{L};p_{R}\right) $ is expressed as
\begin{align}
p =\frac{1}{\sqrt{2}}\left(n+mE; n-mE^{t} \right) e^{-t},
\end{align}
where $Z=\left(m;n \right) =\left(m^{1},\cdots,m^{d} ; n_{1},\cdots,n_{d}\right)$.
The free spectrum of a string is given by the Hamiltonian
\begin{align}
H\sim \frac{1}{2}\left( p_{L}^{2}+p_{R}^{2} \right) =\frac{1}{2}Z\mathcal{M}(E)Z^{t},
\end{align}
where the part of the oscillators is omitted and a $2d\times 2d$ matrix $\mathcal{M}$ is defined as
\begin{align}
\mathcal{M}(E)=\mathcal{E}(e,B)\mathcal{E}^t(e,B)=\left(
\begin{array}{cc}
G-BG^{-1}B & BG^{-1} \\
-G^{-1}B & G^{-1} 
\end{array}
\right).
\end{align}
The T-duality group $O(d,d,\mathbb{Z})$ of the toroidal model acts on $p$ as
\begin{align}\label{O(d,d,Z)}
p\to p'=Zg\mathcal{E}(e,B)\tilde{\mathcal{E}}_{0},~~~~g=\left(
\begin{array}{cc}
a & b \\
c & d 
\end{array}
\right)\in O(d,d,\mathbb{Z}),
\end{align}
where $a,~b,~c,~d$ are $d\times d$ integer matrices that satisfy the following relations:
\begin{align}
a^{t}c+c^{t}a=0,~~b^{t}d+d^{t}b=0,~~a^{t}d+c^{t}b=\boldsymbol{1}_{d}.
\end{align}
We see that the $O(d,d,\mathbb{Z})$ transformation is an automorphism of the free spectrum since the Hamiltonian $H$ transforms to $H'=Zg\mathcal{M}(E)g^{t}Z^{t}$ which gives a point in the same space of states.
The transformation \eqref{O(d,d,Z)} can be interpreted as that of the background matrix $E$ which acts as
\begin{align}\label{fractional linear}
E\to E'=\left( aE+b\right) \left( cE+d\right) ^{-1}.
\end{align}

Let us apply the above discussion about the T-duality group of the non-supersymmetric models to the type II case. 
The non-supersymmetric models are classified by the possible choices of $\hat{Z}=\left(\hat{m};\hat{n} \right)$ with each slot taking 0 or 1 and satisfying $\hat{m}\hat{n}^t=0$ mod 2, except for $\hat{Z}=\left(0^{d};0^{d} \right)$.
The T-duality group \eqref{non-susy duality} in the type II case is written as
\begin{align}\label{type II duality}
D_{(\hat{Z})}(d,d)=\left\lbrace\left.  g=\left(
\begin{array}{cc}
a & b \\
c & d 
\end{array}
\right)\in O(d,d,\mathbb{Z}) \right|~\left(\hat{m}a+\hat{n}c; \hat{m}b+\hat{n}d\right) =\left(\hat{m};\hat{n}\right)~~\text{(mod 2)} \right\rbrace.
\end{align}

\subsubsection{Specific elements of $D_{(\hat{Z})}(d,d)$}

Let us focus on well-known elements of $O(d,d,\mathbb{Z})$ and identify which elements survive in the non-supersymmetric models as symmetries.

\begin{itemize}
	\item Basis change of the compactification lattice:
	\begin{align}
	g_{e}(K)=\left(\begin{array}{cc}
	K & 0 \\
	0 & K^{-t} 
	\end{array}
	\right),~~~~K\in GL\left( d,\mathbb{Z}\right) .
	\end{align}
	In order for $g_{e}(K)$ to be in $D_{(\hat{Z})}(d,d)$, $K$ needs to satisfy $\left(\hat{m}K,\hat{n}K^{-t}\right) =\left(\hat{m},\hat{n}\right) $ mod 2.
	
	\item Integer theta-parameter shift of $B$-field:
	\begin{align}
	g_{B}(\Theta)=\left(\begin{array}{cc}
	\boldsymbol{1}_{d} & \Theta \\
	0 & \boldsymbol{1}_{d}
	\end{array}
	\right) ,~~~~\Theta^{t}=-\Theta\in M_{d\times d}(\mathbb{Z}).
	\end{align}
	From \eqref{type II duality}, the non-supersymmetric model with the choice $\hat{Z}=\left(\hat{m};\hat{n} \right) $ is invariant under the shifts $B_{ij}\to B_{ij}+\Theta_{ij}$ with shift parameters satisfying $\hat{m}\Theta =0$ mod 2.
	
	\item Factorized duality and inversion:
	\begin{align}
	g_{D_{i}}=\left(\begin{array}{cc}
	\boldsymbol{1}_{d}-e_{i} & e_{i} \\
	e_{i} & \boldsymbol{1}_{d}-e_{i}
	\end{array}
	\right),
	\end{align}
	where $e_{i}$ is a $d\times d$ matrix whose components are zero, except for the $ii$ one taking 1. The condition for $g_{D_{i}}$ to be in $D_{(\hat{Z})}(d,d)$ is
	\begin{align}
	\left(\hat{m},\hat{n} \right) =\left(\hat{m}-\hat{m}e_{i}+\hat{n}e_{i},\hat{n}-\hat{n}e_{i}+\hat{m}e_{i} \right)~(\text{mod 2}).
	\end{align}
	Thus, the non-supersymmetric model with the choice $\hat{Z}=\left(\hat{m};\hat{n} \right)$ satisfying $\hat{m}^{i}=\hat{n}_{i}$ is invariant under the $i$-th factorized duality $g_{D_{i}}$. The inversion $g_{D}$ of the background matrix $E$, which is generated by the products of the factorized dualities
	\begin{align}
	g_{D}=\prod_{i=1}^{d}g_{D_{i}}=\left(\begin{array}{cc}
	0& \boldsymbol{1}_{d} \\
	\boldsymbol{1}_{d}& 0
	\end{array}
	\right),
	\end{align}
	is a symmetry only in the non-supersymmetric model with the choice $\hat{Z}=\left(1^{d};1^{d} \right)$.
	
	\item Integer theta-parameter shift of dual $B$-field:
	\begin{align}
	g_{\tilde{B}}(\tilde{\Theta})=g_{D}g_{B}(\tilde{\Theta})g_{D}= \left(\begin{array}{cc}
	\boldsymbol{1}_{d} &0\\
	\beta & \boldsymbol{1}_{d}
	\end{array}
	\right) ,~~~~\tilde{\Theta}^{t}=-\tilde{\Theta}\in M_{d\times d}(\mathbb{Z}).
	\end{align}
	The non-supersymmetric model with the choice $\hat{Z}=\left(\hat{m};\hat{n} \right) $ is invariant under the shifts with parameters satisfying $\hat{n}\tilde{\Theta}=0$ mod 2.
\end{itemize}  

The first two elements are called geometric ones. Indeed one can check $\mathcal{E}(e,B)=g_{e}(e)g_{B}(B)$, and hence any generalized vierbeins are obtained by starting from $\mathcal{E}(\boldsymbol{1}_{d},0)=\boldsymbol{1}_{2d}$ and acting $g_{e}$ and $g_{B}$. On the other hand, $g_{D_{i}}$, $g_{D}$ and $g_{\tilde{B}}$ are known as non-geometric elements.

The simplest example is the $d=1$ case in which we have two inequivalent choices $\hat{Z}=(1,0),(0,1)$\footnote{The condition \eqref{delta2integer} prohibits $(\hat{m},\hat{n})=(1,1)$. }. There is only one non-trivial element in $O(1,1,\mathbb{Z})$, that is, the factorized duality $g_{D_{1}}$. The factorized duality cannot be a symmetry of the non-supersymmetric models since neither of the choices of $\hat{Z}$ satisfies $\hat{m}^{1}=\hat{n}_{1}$ mod 2. Rather than that, acting $g_{D_{1}}$ on either of the models produces the other model. These models interpolate between two different 10D endpoint string models with a volume parameter $R$, and the factorized duality interchanges two of the endpoint models.

\subsubsection{$d=2$ in the type II case}

One of the simple and interesting examples is the $d=2$ case. We can change the basis of the moduli space such that the duality symmetry $O(2,2,\mathbb{Z})$ is decomposed into $PSL(2,\mathbb{Z})\times PSL(2,\mathbb{Z})$.
To do this, we define two complex parameters $\tau$ and $\rho$ by combining the four real parameters $G_{11},~G_{22},~G_{12},~B_{12}$ as follows:
\begin{subequations}\label{modular basis}
	\begin{align}
	\tau&=\tau_{1}+i\tau_{2}=\frac{G_{12}}{G_{22}}+i\frac{\sqrt{G}}{G_{22}},\\
	\rho&=\rho_{1}+i\rho_{2}=B_{12}+i\sqrt{G},
	\end{align}
\end{subequations}
where $G=G_{11}G_{22}-G_{12}^{2}$. Then the four real momenta $p_{Li}$, $p_{Ri}$ are expressed as two complex ones:
\begin{subequations}
	\begin{align}
	\left| p_{L}\right| &=\frac{1}{\sqrt{2\tau_{2}\rho_{2}}}\left|  \left(n_{1}-\tau n_{2}\right)  -\rho\left(m_{2}+\tau m_{1} \right) \right| ,  \\
	\left| p_{R}\right| &=\frac{1}{\sqrt{2\tau_{2}\rho_{2}}}\left|  \left(n_{1}-\tau n_{2}\right)  -\bar{\rho}\left(m_{2}+\tau m_{1} \right)\right| .
	\end{align}
\end{subequations}
In the toroidal models, one can find two modular symmetries which act on the complex structure $\tau$ and the K\"{a}hler structure $\rho$ respectively.
\begin{align}
\label{modular group 1}
g_{\tau}(\gamma):~\left( \tau,\rho\right)\to  \left( \frac{a\tau+b}{c\tau+d},\rho\right) ,\\
\label{modular group 2}
g_{\rho}(\gamma):~\left( \tau,\rho\right)\to  \left( \tau,\frac{a\rho+b}{c\rho+d}\right),
\end{align}
where $\gamma=\left(
\begin{array}{cc}
a & b \\
c & d
\end{array}
\right)\in PSL(2,\mathbb{Z})$.  Besides the modular groups, there are some discrete symmetries. One of them is the interchange of the complex and K\"ahler structures
\begin{align}
g_{D_{2}}:~(\tau,\rho)\to (\rho,\tau),
\end{align}
which corresponds to the factorized duality for the $X^{2}$-direction. The factorized duality for the $X^{1}$-direction is obtained by the following transformation:
\begin{align}
g_{D_{1}}:~(\tau,\rho)\to \left( -\frac{1}{\rho},-\frac{1}{\tau}\right).
\end{align}
The interchange of the basis $X^{1}\leftrightarrow X^{2}$ is 
\begin{align}
g_{S^{2}}:~(\tau,\rho)\to \left( \frac{1}{\bar{\tau}},-\bar{\rho}\right).
\end{align}
The others are the reflection $X_{2}\to -X_{2}$ and the world sheet parity $p_{L}\leftrightarrow p_{R}$, which are respectively expressed as the following transformations:
\begin{align}
g_{R}:~(\tau,\rho)\to (-\bar{\tau},-\bar{\rho}),~~~~g_{W}:~(\tau,\rho)\to (\tau,-\bar{\rho}).
\end{align} 
All the elements that we present above are not independent. 
In fact, we can pick up the four elements $g_{\tau}(\gamma_{T})$, $g_{\tau}(\gamma_{S})$, $g_{W}$ and $g_{D_{2}}$ as a minimum set of the generators. Here $\gamma_{T}$ and $\gamma_{S}$ are matrices generating a modular group:
\begin{align}
\gamma_{T}=\left(
\begin{array}{cc}
1 & 1 \\
0 & 1
\end{array}
\right),~~~~\gamma_{S}=\left(
\begin{array}{cc}
0 & 1 \\
-1 & 0
\end{array}
\right).
\end{align}
The other elements are obtained by the combinations of the generators. For instance, the modular group \eqref{modular group 2} acting on $\rho$ is generated by 
\begin{align}
g_{\rho}(\gamma_{T})=g_{D_{2}}g_{\tau}(\gamma_{T})g_{D_{2}},~~~~g_{\rho}(\gamma_{S})=g_{D_{2}}g_{\tau}(\gamma_{S})g_{D_{2}}.
\end{align}
The $\mathbb{Z}_{2}$ elements $g_{R}$, $g_{S^{2}}$ and $g_{D_{1}}$ can be also expressed as the products of the generators, as shown in Table 3.
\begin{table}[t]
	\begin{center}
		\begin{tabular}{|c|c|c|c|c|} \hline
			$g_{\rho}(\gamma_{T})$ & $g_{\rho}(\gamma_{S})$ & $g_{R}$ & $g_{S^{2}}$ & $g_{D_{1}}$ \\\hline
			$g_{D_{2}}g_{\tau}(\gamma_{T})g_{D_{2}}$& $g_{D_{2}}g_{\tau}(\gamma_{S})g_{D_{2}}$ & $g_{D_{2}}g_{W}g_{D_{2}}g_{W}$ &  $g_{R}g_{\tau}(\gamma_{S})$ &  $g_{S^{2}}g_{D_{2}}g_{S^{2}}$ \\ 			
			\hline
		\end{tabular}
		\caption{The elements $g_{\tau}(\gamma_{T})$, $g_{\tau}(\gamma_{T})$, $g_{W}$ and $g_{D_{2}}$ generate the T-duality group. This table lists the products of the generators which give $g_{\rho}(\gamma)$, $g_R$, $g_{S^{2}}$ and $g_{D_{1}}$.}
	\end{center}
\end{table}	

Transformations of $\left( \tau,\rho\right)$ can be regarded as those of $Z=\left(m;n \right) $. Under $g_{\tau}(\gamma)$, $g_{D_{2}}$ and $g_{W}$, for instance, $Z$ transforms as
\begin{align}\label{action on Z}
g_{\tau}(\gamma)&:~Z\to ZM_{\tau}\left(\gamma\right) ,\\
g_{D_{2}}&:~Z\to ZM_{D_{2}},\\
g_{W}&:~Z\to ZM_{W},
\end{align}
where $M_{\tau}\left(\gamma\right)$, $M_{D_{2}}$ and $M_{W}$ are $4\times 4$ matrices defined as
\begin{align}
M_{\tau}\left(\gamma\right)=\left(
\begin{array}{cc}
\gamma & 0 \\
0 & \gamma^{-t}
\end{array}
\right),~~~~
M_{D_{2}}=\left(
\begin{array}{cc}
e_{1} & e_{2} \\
e_{2} & e_{1}
\end{array}
\right),~~~~
M_{W}=\left(
\begin{array}{cc}
-\boldsymbol{1}_{2} & 0 \\
0 & \boldsymbol{1}_{2}
\end{array}
\right),
\end{align}
where $e_{1}=diag(1,0)$ and $e_{2}=diag(0,1)$. The representation matrices of the other elements are expressed as the products of $M_{\tau}\left(\gamma\right)$, $M_{W}$ and $M_{D_{2}}$. As shown in Table 3, for instance,  the representation matrices of $g_{\rho}(\gamma)$ and $g_{R}$ are given by
\begin{align}
M_{\rho}(\gamma)=M_{D_{2}}M_{\tau}(\gamma)M_{D_{2}},~~~~M_{R}=M_{D_{2}}M_{W}M_{D_{2}}M_{W}.
\end{align}

\begin{table}[t]
	\begin{center}
		\begin{tabular}{|c||c|c|c|c|c|} \hline
			$\left(\hat{m};\hat{n} \right) $ & $g_{\tau}(\gamma)$ & $g_{\rho}(\gamma)$ & $~g_{D_{2}}~$ & $~g_{D_{1}}~$& $~g_{S^{2}}~$ \\\hline
			$\left(1,0;0,0\right)$& $\gamma\in\Gamma^{1}(2)$ & $\gamma\in\Gamma^{1}(2)$ &$g_{D_{2}}$& ---& --- \\
			$\left(0,1;0,0\right)$& $\gamma\in\Gamma_{1}(2)$ & $\gamma\in\Gamma^{1}(2)$ & --- & $g_{D_{1}}$& --- \\
			$\left(0,0;1,0\right)$& $\gamma\in\Gamma_{1}(2)$ & $\gamma\in\Gamma_{1}(2)$ & $g_{D_{2}}$& ---& --- \\
			$\left(0,0;0,1\right)$& $\gamma\in\Gamma^{1}(2)$ & $\gamma\in\Gamma_{1}(2)$ & ---&$g_{D_{1}}$& --- \\
			$\left(1,0;0,1\right)$& $\gamma\in\Gamma^{1}(2)$ & $\gamma\in\Gamma_{\vartheta}$ & --- & ---& ---\\
			$\left(0,1;1,0\right)$& $\gamma\in\Gamma_{1}(2)$ & $\gamma\in\Gamma_{\vartheta}$ & --- & ---& ---\\
			$\left(1,1;0,0\right)$& $\gamma\in\Gamma_{\vartheta}$ & $\gamma\in\Gamma^{1}(2)$ & --- & ---& $g_{S^{2}}$\\
			$\left(0,0;1,1\right)$& $\gamma\in\Gamma_{\vartheta}$ & $\gamma\in\Gamma_{1}(2)$ & --- & ---& $g_{S^{2}}$\\
			$\left(1,1;1,1\right)$& $\gamma\in\Gamma_{\vartheta}$ & $\gamma\in\Gamma_{\vartheta}$ & $g_{D_{2}}$ & $g_{D_{1}}$ & $g_{S^{2}}$\\
			\hline
		\end{tabular}
		\caption{The elements of $D_{(\hat{Z})}(2,2)$ which depend on the choice of $\hat{Z}$ are shown. }
	\end{center}
\end{table}	

Let us study the T-duality group $D_{(\hat{Z})}(2,2)$ of the non-supersymmetric model on the basis given in \eqref{modular basis}.
There are nine possible choices of $\hat{Z}$ with $d=2$: $\hat{Z}=\left(\underline{1,0};0,0\right)$, $\left(0,0;\underline{1,0}\right)$, $\left(1,0;0,1 \right)$, $\left(0,1;1,0 \right)$, $\left(1,1;0,0 \right)$, $\left(0,0;1,1 \right)$, $\left(1,1;1,1 \right)$. Here the underline indicates the permutation of the components. We can identify the elements of $D_{(\hat{Z})}(2,2)$ from the actions of $g$ on $\hat{Z}$. For the modular group \eqref{modular group 1}, $g_{\tau}(\gamma)$ is in $D_{(\hat{Z})}(2,2)$ if $\gamma$ satisfies
\begin{align}
\left(\hat{m};\hat{n} \right)&=\left(\hat{m}\gamma;\hat{n}\gamma^{-t} \right) ~(\text{mod 2}).
\end{align}
The other elements of $D_{(\hat{Z})}(2,2)$ can be identified in the same way by using the corresponding representation matrices. Note that the reflection $g_{R}$ and the world-sheet parity $g_{W}$ are in $D_{(\hat{Z})}(2,2)$ whatever the choice of $\hat{Z}$ is since the representation matrices are diagonal.
The specific elements of $D_{(\hat{Z})}(2,2)$ are shown in Table 4.
Here $\Gamma_{1}(n)$ and $\Gamma^{1}(n)$ are the Hecke congruence subgroups of the modular group
\begin{align}
\Gamma_{1}(n)&=\left\lbrace \left. \left(\begin{array}{cc}
a & b \\
c & d
\end{array}
\right) \in PSL(2,\mathbb{Z})~\right| ~a,d=1,~c=0~(\text{mod}~n) \right\rbrace,\\
\Gamma^{1}(n)&=\left\lbrace \left. \left(\begin{array}{cc}
a & b \\
c & d
\end{array}
\right) \in PSL(2,\mathbb{Z})~\right| ~a,d=1,~b=0~(\text{mod}~n) \right\rbrace,
\end{align}
and $\Gamma_{\vartheta}$ is the theta subgroup
\begin{align}
\Gamma_{\vartheta}&=\left\lbrace \left. \left(\begin{array}{cc}
a & b \\
c & d
\end{array}
\right) \in PSL(2,\mathbb{Z})~\right| ~ac=0,~bd=0~(\text{mod}~2) \right\rbrace.
\end{align}

We can see that the transitions among the non-supersymmetric models are induced by acting $g$ on the models in which $g$ is not a symmetry. Focusing on $\gamma_{T}$ and $\gamma_{T}^{-t}$, we notice
\begin{align}
\gamma_{T}\notin \Gamma^{1}(2),\Gamma_{\vartheta},~~~~\gamma_{T}^{-t}\notin \Gamma_{1}(2),\Gamma_{\vartheta}.
\end{align}
For example, starting from the model with the choice $\hat{Z}=\left(1,1;1,1 \right)$, we can obtain all of the other models by acting on $g_{\tau}(\gamma_{T})$, $g_{\tau}(\gamma_{T}^{-t})$, $g_{\rho}(\gamma_{T})$ or $g_{\rho}(\gamma_{T}^{-t})$ successively and appropriately (see Fig. 1).

\begin{figure}[t]
	\begin{center}
		\includegraphics[width=15cm]{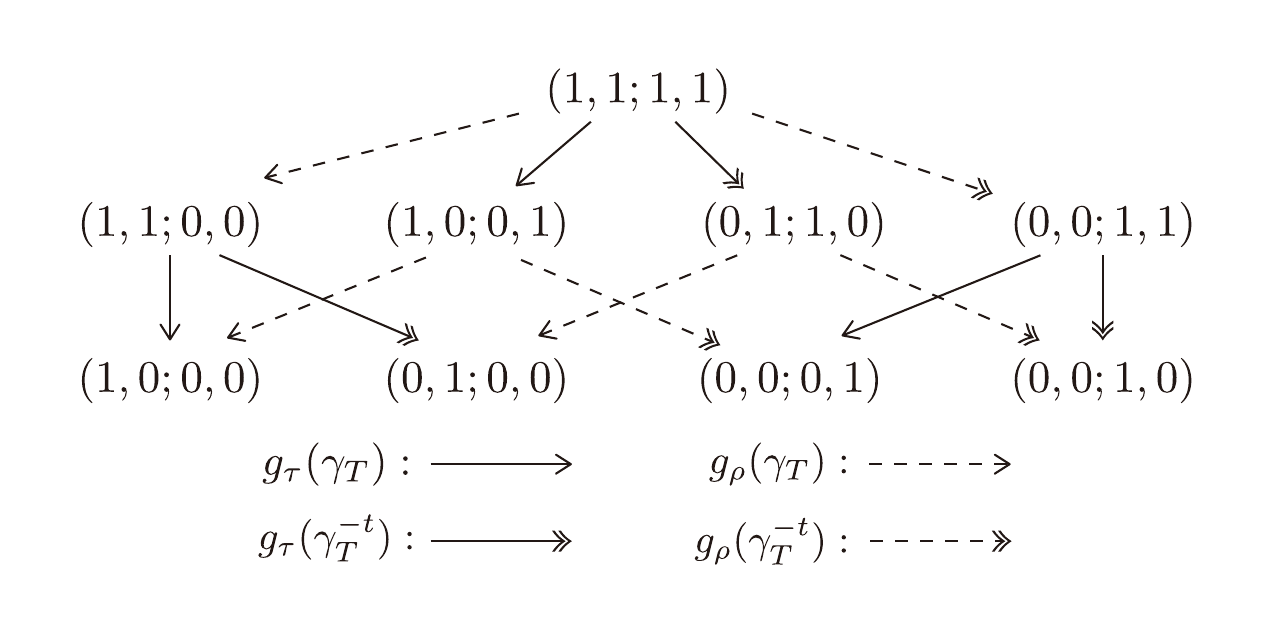}
	\end{center}
	\label{fig:one}
	\vspace{-10mm}
	\caption{An example of the transitions among the non-supersymmetric type II models with $d=2$ is shown.}
\end{figure}

\subsection{T-duality in the heterotic case}\label{T-duality heterotic}

In the heterotic models $d$-dimensional toroidal compactified, there are $(16+d)\times d$ moduli: a metric $G=ee^t$ of the compactification lattice, an anti-symmetric two-form $B$ and Wilson lines $A$.
We can choose a Narain metric as
\begin{align}
J =\left(\begin{array}{ccc}
g_{16} & 0 & 0\\
0 & 0 & \boldsymbol{1}_{d}\\
0 & \boldsymbol{1}_{d} & 0
\end{array}
\right),~~~g_{16}=\alpha_{16}\alpha_{16}^{t}
\end{align}
where $\alpha_{16}$ denotes a set of the basis of a 16-dimensional even self-dual Euclidean lattice $\Gamma^{16}$. An internal momentum $p\in \Gamma^{16+d,d}$ is then expressed as
\begin{align}
p=Z\mathcal{E}(e,B,A) \tilde{\mathcal{E}}_{0},
\end{align}
where $Z=\left(q,m,n \right)\in \mathbb{Z}^{16}\times \mathbb{Z}^d\times \mathbb{Z}^d$ and
\begin{align}\label{generalized vierbein hetero}
\mathcal{E}(e,B,A) =\left(\begin{array}{ccc}
\boldsymbol{1}_{16} & 0 & \alpha_{16}A^{t}e^{-t}\\
-A \alpha_{16}^{-1}&  e & -C^te^{-t}\\
0 & 0 & e^{-t}
\end{array}
\right),~~~
\tilde{\mathcal{E}}_{0} =\left(\begin{array}{ccc}
\alpha_{16} & 0 & 0\\
0 & \frac{1}{\sqrt{2}}\boldsymbol{1}_{d} & -\frac{1}{\sqrt{2}}\boldsymbol{1}_{d}\\
0 & \frac{1}{\sqrt{2}}\boldsymbol{1}_{d} & \frac{1}{\sqrt{2}}\boldsymbol{1}_{d}
\end{array}
\right),
~~~C=B+\frac{1}{2}AA^t.
\end{align}
Writing down $p=(\ell_{L},p_{R},p_{L})$ explicitly, we get
\begin{subequations}\label{momenta(explicit)}
	\begin{align}
	\ell_{L}&=  \pi-mA,\\
	p_{L}&=\frac{1}{\sqrt{2}}\left[   \pi A^{t}+m \left( G-C^t\right) +n \right] e^{-t} ,\\
	p_{R}&=\frac{1}{\sqrt{2}}\left[  \pi A^{t}-m \left( G+C^t\right) +n\right]  e^{-t},
	\end{align}
\end{subequations}
where $\pi=q\alpha_{16}$ lives in $\Gamma^{16}$. 
One can check that $\tilde{\mathcal{E}}_{0}$ and $\mathcal{E}$ satisfy $\tilde{\mathcal{E}}_{0} \eta \tilde{\mathcal{E}}_{0}^{t}=J$, $\mathcal{E} J\mathcal{E}^{t}=J$ and the inner product is independent of the moduli:
\begin{align}
p_{1}\cdot p_{2}= p_{1}\eta p_{2}^{t}=Z_{1} J Z_{2}^{t}=\pi_{1}\pi_{2}^{t}+m_{1}n_{2}^t+n_{1}m_{2}^t.
\end{align}
The T-duality group of the toroidal model acts on $p$ as
\begin{align}
p\to p'=Zg\mathcal{E}(e,B,A) \tilde{\mathcal{E}}_{0},~~~~g\in O(16+d,d,\mathbb{Z}),
\end{align}
where $g$ is a $(16+2d)\times (16+2d)$ integer matrix that satisfies $gJg^{t}=J$.

Choosing a certain set of integers $\hat{Z}=\left(\hat{q},\hat{m},\hat{n} \right)$ that satisfies
\begin{align}\label{integer condition}
\left| \hat{\pi}\right| ^2+2\hat{m}\hat{n}^{t}=0~(\text{mod 4}),
\end{align}
where $\hat{\pi}=\hat{q}\alpha_{16}$ and $|\hat{\pi}|^{2}=\hat{\pi}\hat{\pi}^{t}$, the shift vector $\delta_{(\hat{Z})}$ is expressed as
\begin{align}
\delta_{(\hat{Z})}=\frac{1}{2}\hat{Z}\mathcal{E}(e,B,A) \tilde{\mathcal{E}}_{0}.
\end{align}
The T-duality group of the non-supersymmetric heterotic model with the choice $\hat{Z}$ is
\begin{align}
D_{(\hat{Z})}\left(16+d,d \right) =\left\lbrace g\in O(16+d,d,\mathbb{Z}) \left| ~\hat{Z}=\hat{Z}g~(\text{mod 2}) \right. \right\rbrace.
\end{align}

\subsubsection{Specific elements of $D_{(\hat{Z})}(16+d,d)$}
Let us see specific elements of $O(16+d,d,\mathbb{Z})$ and identify the congruence conditions that are imposed on the elements of $D_{(\hat{Z})}(16+d,d)$.
\begin{itemize}
	\item Basis change of the compactification lattice:
	\begin{align}
	g_{e}(K)=\left(\begin{array}{ccc}
	\boldsymbol{1}_{16}&0&0\\
	0&K & 0 \\
	0&0 & K^{-t} 
	\end{array}
	\right),~~~~K\in GL\left( d,\mathbb{Z}\right) .
	\end{align}
	The elements $g_{K}$ of $D_{(\hat{Z})}(16+d,d)$ must satisfy $\left( \hat{m}K, \hat{n}K^{-t}\right)= \left( \hat{m}, \hat{n}\right)$ mod 2.
	
	\item Basis change of the gauge lattice:
	\begin{align}
	g_{\Gamma^{16}}(W)=\left(\begin{array}{ccc}
	\alpha_{16}W\alpha_{16}^{-1}&0&0\\
	0&\boldsymbol{1}_{d} & 0 \\
	0&0 & \boldsymbol{1}_{d} 
	\end{array}
	\right),~~~~W\in O\left( 16,\mathbb{Z}\right) .
	\end{align}
	Acting $g_{\Gamma^{16}}(W)$ on $\mathcal{E}$, the Wilson lines transform as $A\to AW^{t}$ while $G$ and $B$ are unchanged. 
	Acting $g_{\Gamma^{16}}(W)$ leads to a change of the basis of $\Gamma^{16}$ as $\pi\to\pi W$ accompanied with the $O(16+d)\times O(d)$ rotation $\left(\ell_{L},p_{L},p_{R} \right) \to \left(\ell_{L}W^{t},p_{L},p_{R} \right)$.
	The condition for $g_{\Gamma^{16}}(W)$ to be in $D_{(\hat{Z})}(16+d,d)$ is $\hat{\pi}W=\hat{\pi}+2\pi_{0}$ for ${}^{\exists}\pi_{0}\in \Gamma^{16}$.
	
	\item Integer theta-parameter shift of $B$-field:
	\begin{align}
	g_{B}(\Theta)=
	\left(\begin{array}{ccc}
	\boldsymbol{1}_{16}&0&0\\
	0&\boldsymbol{1}_{d}&\Theta  \\
	0&0 & \boldsymbol{1}_{d}
	\end{array}
	\right) ,~~~~\Theta^{t}=-\Theta\in M_{d\times d}(\mathbb{Z}).
	\end{align}
	If a shift parameter $\Theta$ satisfies $\hat{m}\Theta=0$ mod 2 then $g_{B}(\Theta)$ is an element of $D_{(\hat{Z})}(16+d,d)$.
	
	\item Wilson line shift:
	\begin{align}
	g_{A}(a)=
	\left(\begin{array}{ccc}
	\boldsymbol{1}_{16}&0&g_{16}a^{t}\\
	-a&\boldsymbol{1}_{d}&-\frac{1}{2}ag_{16}a^{t}\\
	0&0 & \boldsymbol{1}_{d}
	\end{array}
	\right) ,~~~~a\in M_{d\times 16}(\mathbb{Z}).
	\end{align}
	Under $g_{A}(a)$, the Wilson lines $A$ and the two-form $B$ are shifted as
	\begin{align}
	A\to A+\pi_{a},~~~~B\to B+\frac{1}{2}\left(A\pi_{a}^{t}-\pi_{a}A^{t} \right),
	\end{align}
	where $\pi_{a}=a\alpha_{16}$. The elements $g_{A}(a)$ of $D_{(\hat{Z})}(16+d,d)$ satisfy both of the following conditions:
	\begin{align}
	\hat{m}a=0~(\text{mod 2}),~~\left( \hat{\pi}-\frac{1}{2}\hat{m}\pi_{a}\right)\pi_{a}^{t}=0~(\text{mod 2}).
	\end{align}
	
	\item Factorized duality and inversion:
	\begin{align}\label{factorized duality in hetero}
	g_{D_{i}}=\left(\begin{array}{ccc}
	\boldsymbol{1}_{16}&0&0\\
	0&\boldsymbol{1}_{d}-e_{i} & e_{i} \\
	0&e_{i} & \boldsymbol{1}_{d}-e_{i}
	\end{array}
	\right),
	\end{align}
	The non-supersymmetric models with the choice $\hat{Z}$ satisfying $\hat{m}^{i}=\hat{n}_{i}$ have the $i$-th factorized duality symmetry $g_{D_{i}}$. The inversion $g_{D}$, which is expressed as
	\begin{align}\label{inversion in hetero}
	g_{D}=\prod_{i=1}^{d}g_{D_{i}}=\left(\begin{array}{ccc}
	\boldsymbol{1}_{16}&0&0\\
	0&0 & \boldsymbol{1}_{d} \\
	0&\boldsymbol{1}_{d} & 0
	\end{array}
	\right),
	\end{align}
	is an element of $D_{(\hat{Z})}(16+d,d)$ with the choice $\hat{Z}$ satisfying $\hat{m}=\hat{n}$ for all directions.
	
	\item Integer theta-parameter shift of dual $B$-field:
	\begin{align}
	g_{\tilde{B}}(\tilde{\Theta})=g_{D}g_{B}(\tilde{\Theta})g_{D}= \left(\begin{array}{ccc}
	\boldsymbol{1}_{16}&0&0\\
	0&\boldsymbol{1}_{d} &0\\
	0&\tilde{\Theta} & \boldsymbol{1}_{d}
	\end{array}
	\right) ,~~~~\tilde{\Theta}^{t}=-\tilde{\Theta}\in M_{d\times d}(\mathbb{Z}).
	\end{align}
	If a shift parameter $\tilde{\Theta}$ satisfies $\hat{n}\tilde{\Theta}=0$ mod 2 then $g_{\tilde{B}}(\tilde{\Theta})$ is an element of $D_{(\hat{Z})}(16+d,d)$.

	\item dual Wilson line shift:
	\begin{align}
	g_{\tilde{A}}\left( \tilde{a}\right) =g_{D}g_{A}(\tilde{a})g_{D}=
	\left(\begin{array}{ccc}
	\boldsymbol{1}_{16}&g_{16}\tilde{a}^{t}&0\\
	0&\boldsymbol{1}_{d}&0\\
	-\tilde{a}&-\frac{1}{2}\tilde{a}g_{16} \tilde{a}^t & \boldsymbol{1}_{d}
	\end{array}
	\right) ,~~~~\tilde{a}\in M_{d\times 16}(\mathbb{Z}).
	\end{align}
	The elements $g_{\tilde{A}}\left( \tilde{a}\right)$ of $D_{(\hat{Z})}(16+d,d)$ satisfy both of the following conditions:
	\begin{align}
	\hat{n}\tilde{a}=0~(\text{mod 2}),~~\left( \hat{\pi}-\frac{1}{2}\hat{n}\pi_{\tilde{a}}\right)\pi_{\tilde{a}}^{t}=0~(\text{mod 2}).
	\end{align}
\end{itemize}  

The first four elements are geometric ones and the last three elements are non-geometric ones. Indeed, one can check $\mathcal{E}(e,B,A)=g_{A}(A\alpha_{16}^{-1})g_{B}(B)g_{e}(e)$ from \eqref{generalized vierbein hetero} as in the type II case.

\subsubsection{$d=1$ in the heterotic case}
Unlike in the type II case, there are a lot of T-duality elements in the heterotic case even if $d=1$. 
The momenta \eqref{momenta(explicit)} with $d=1$ are written as
\begin{subequations}\label{momenta_d=1}
	\begin{align}
	\ell_{L}&=  \pi-mA,\\
	p_{L}&=\frac{1}{\sqrt{2}R}\left[   \pi A^{t}+m \left( R^{2}-\frac{1}{2}\left| A\right|^{2}\right) +n \right] ,\\
	p_{R}&=\frac{1}{\sqrt{2}R}\left[  \pi A^{t}-m \left( R^{2}+\frac{1}{2}\left| A\right|^{2} \right) +n\right] ,
	\end{align}
\end{subequations}
where $R$ is a radius of a circle and $|A|^{2}=AA^{t}$.
From the condition \eqref{integer condition}, we can classify the 9D non-supersymmetric heterotic models into the following four classes;
\begin{enumerate}
	\item $\left|\hat{\pi} \right|^2=0~(\text{mod 4}),~~\left(\hat{m},\hat{n} \right) =(0,0)$;\\
	In this class, $\Gamma_{\pm}^{17,1}$ and $\Gamma_{\pm}^{17,1}+\delta$ are written as the following sets:
	\begin{align}
	\label{class1Gammapm}
	&\Gamma_{\pm}^{17,1}=\left\lbrace p=Z\tilde{\mathcal{E}}\left|  \left(\pi,m,n \right)\in \left(\Gamma^{16}_{\pm},\mathbb{Z},\mathbb{Z} \right)  \right.  \right\rbrace,\\
	\label{class1Gammapmdelta}
	&\Gamma_{\pm}^{17,1}+\delta=\left\lbrace p=Z\tilde{\mathcal{E}}\left|\left(\pi,m,n \right)\in \left(\Gamma^{16}_{\pm}+\frac{\hat{\pi}}{2},\mathbb{Z},\mathbb{Z} \right) \right.  \right\rbrace,
	\end{align}
	where $\Gamma_{+}^{16}(\hat{\pi})$ and $\Gamma_{-}^{16}(\hat{\pi})$ are defined as
	\begin{align}\label{Gamma16pm}
	\Gamma_{+}^{16}(\hat{\pi})=\left\lbrace \left.  \pi\in \Gamma^{16}~\right|  \hat{\pi}\cdot \pi\in 2\mathbb{Z} \right\rbrace,~~~~~
	\Gamma_{-}^{16}(\hat{\pi})=\left\lbrace \left.  \pi\in \Gamma^{16}~\right|  \hat{\pi}\cdot \pi\in 2\mathbb{Z}+1 \right\rbrace.
	\end{align}
	The non-supersymmetric models in this class are obtained by compactifying the 10D non-supersymmetric heterotic models shown in Tables 1 and 2 on a circle. To see this, let us study the behaviors in the endpoint limits ($R\to \infty$ and $R\to 0$) with $A=0$. Note that the states with $m=0$ ($n=0$) only contribute as $R\to\infty$ ($R\to 0$). We then find the behaviors of $Z_{\Gamma_{\pm}^{17,1}}$ and $Z_{\Gamma_{\pm}^{17,1}+\delta}$ from \eqref{class1Gammapm} and \eqref{class1Gammapmdelta}:
	\begin{align}
	&Z_{\Gamma_{\pm}^{17,1}}\to \frac{R}{\sqrt{\tau_{2}}}\left(\eta\bar{\eta} \right)^{-1} Z_{\Gamma_{\pm}^{16}},~~~Z_{\Gamma_{\pm}^{17,1}+\delta}\to \frac{R}{\sqrt{\tau_{2}}}\left(\eta\bar{\eta} \right)^{-1}Z_{\Gamma_{\pm}^{16}+\frac{\hat{\pi}}{2}},~~~~~(R\to \infty),\\
	&Z_{\Gamma_{\pm}^{17,1}}\to \frac{1}{R\sqrt{\tau_{2}}}\left(\eta\bar{\eta} \right)^{-1}Z_{\Gamma_{\pm}^{16}},~~~Z_{\Gamma_{\pm}^{17,1}+\delta}\to \frac{1}{R\sqrt{\tau_{2}}}\left(\eta\bar{\eta} \right)^{-1}Z_{\Gamma_{\pm}^{16}+\frac{\hat{\pi}}{2}},~~~~~(R\to 0),
	\end{align}
	where $Z_{\Gamma_{\pm}^{16}}$ and $Z_{\Gamma_{\pm}^{16}+\frac{\hat{\pi}}{2}}$ are defined as
	\begin{align}
	Z_{\Gamma_{\pm}^{16}}=\eta^{-16}\sum_{\pi\in\Gamma^{16}_{\pm}}q^{\frac{1}{2}|\pi|^2},~~~~Z_{\Gamma_{\pm}^{16}+\frac{\hat{\pi}}{2}}=\eta^{-16}\sum_{\pi\in\Gamma^{16}_{\pm}}q^{\frac{1}{2}\left| \pi+\frac{\hat{\pi}}{2}\right| ^2}.
	\end{align}
	Both of the endpoint limits in this class then give the same 10D non-supersymmetric model constructed by using the shift-vector $\delta=\hat{\pi}/2$.
	
	\item $\left|\hat{\pi} \right|^2=0~(\text{mod 4}),~~\left(\hat{m},\hat{n} \right) =(1,0)$;\\
	In this class, $\Gamma_{\pm}^{17,1}$ and $\Gamma_{\pm}^{17,1}+\delta$ are written as 
	\begin{align}
	\label{class2Gammapm}
	&\Gamma_{\pm}^{17,1}=\left\lbrace p=Z\tilde{\mathcal{E}}\left| \left(\pi,m,n \right)\in\left( \Gamma^{16}_{\pm},\mathbb{Z},2\mathbb{Z}\right) \right.  \right\rbrace+
	\left\lbrace p=Z\tilde{\mathcal{E}}\left| \left(\pi,m,n \right)\in\left( \Gamma^{16}_{\mp},\mathbb{Z},2\mathbb{Z}+1\right)  \right.  \right\rbrace,\\
	\label{class2Gammapmdelta}
	&\Gamma_{\pm}^{17,1}+\delta=\left\lbrace p=Z\tilde{\mathcal{E}}\left| \left(\pi,m,n \right)\in\left( \Gamma^{16}_{\pm}+\frac{\hat{\pi}}{2},\mathbb{Z}+\frac{1}{2},2\mathbb{Z}\right)  \right.  \right\rbrace
	\nonumber\\&~~~~~~~~~~~~~~~
	+\left\lbrace p=Z\tilde{\mathcal{E}} \left|  \left(\pi,m,n \right)\in\left( \Gamma^{16}_{\mp}+\frac{\hat{\pi}}{2},\mathbb{Z}+\frac{1}{2},2\mathbb{Z}+1\right) \right.  \right\rbrace.
	\end{align}
	From \eqref{class2Gammapm} and \eqref{class2Gammapmdelta}, the behaviors of $Z_{\Gamma_{\pm}^{17,1}}$ and $Z_{\Gamma_{\pm}^{17,1}+\delta}$ in the endpoint limits are
	\begin{align}
	&Z_{\Gamma_{\pm}^{17,1}}\to \frac{R}{\sqrt{\tau_{2}}}\left(\eta\bar{\eta} \right)^{-1}Z_{\Gamma^{16}},~~~Z_{\Gamma_{\pm}^{17,1}+\delta}\to 0,~~~~~(R\to \infty),\\
	&Z_{\Gamma_{\pm}^{17,1}}\to \frac{1}{R\sqrt{\tau_{2}}}\left(\eta\bar{\eta} \right)^{-1}Z_{\Gamma_{\pm}^{16}},~~~Z_{\Gamma_{\pm}^{17,1}+\delta}\to \frac{1}{R\sqrt{\tau_{2}}}\left(\eta\bar{\eta} \right)^{-1}Z_{\Gamma_{\pm}^{16}+\frac{\hat{\pi}}{2}},~~~~~(R\to 0).
	\end{align}
	These behaviors imply that the supersymmetry is asymptotically restoring in the limit $R\to\infty$ while the 10D non-supersymmetric heterotic models are produced in the limit $R\to0$.

	\item $\left|\hat{\pi} \right|^2=0~(\text{mod 4}),~~\left(\hat{m},\hat{n} \right) =(0,1)$;\\
	In this class, $\Gamma_{\pm}^{17,1}$ and $\Gamma_{\pm}^{17,1}+\delta$ are written as 
	\begin{align}
	\label{class3Gammapm}
	&\Gamma_{\pm}^{17,1}=\left\lbrace p=Z\tilde{\mathcal{E}}\left| \left(\pi,m,n \right)\in\left( \Gamma^{16}_{\pm},2\mathbb{Z},\mathbb{Z}\right) \right.  \right\rbrace+
	\left\lbrace p=Z\tilde{\mathcal{E}}\left| \left(\pi,m,n \right)\in\left( \Gamma^{16}_{\mp},2\mathbb{Z}+1,\mathbb{Z}\right)  \right.  \right\rbrace,\\
	\label{class3Gammapmdelta}
	&\Gamma_{\pm}^{17,1}+\delta=\left\lbrace p=Z\tilde{\mathcal{E}}\left| \left(\pi,m,n \right)\in\left( \Gamma^{16}_{\pm}+\frac{\hat{\pi}}{2},2\mathbb{Z},\mathbb{Z}+\frac{1}{2}\right)  \right.  \right\rbrace
	\nonumber\\&~~~~~~~~~~~~~~~
	+\left\lbrace p=Z\tilde{\mathcal{E}} \left|  \left(\pi,m,n \right)\in\left( \Gamma^{16}_{\mp}+\frac{\hat{\pi}}{2},2\mathbb{Z}+1,\mathbb{Z}+\frac{1}{2}\right) \right.  \right\rbrace.
	\end{align}
	The behaviors of $Z_{\Gamma_{\pm}^{17,1}}$ and $Z_{\Gamma_{\pm}^{17,1}+\delta}$ in the endpoint limits are 
	\begin{align}
	&Z_{\Gamma_{\pm}^{17,1}}\to \frac{R}{\sqrt{\tau_{2}}}\left(\eta\bar{\eta} \right)^{-1}Z_{\Gamma^{16}_{\pm}},
	~~~Z_{\Gamma_{\pm}^{17,1}+\delta}\to \frac{R}{\sqrt{\tau_{2}}}\left(\eta\bar{\eta} \right)^{-1}Z_{\Gamma^{16}_{\pm}+\frac{\hat{\pi}}{2}},~~~~~(R\to \infty),\\
	&Z_{\Gamma_{\pm}^{17,1}}\to \frac{1}{R\sqrt{\tau_{2}}}\left(\eta\bar{\eta} \right)^{-1}Z_{\Gamma^{16}},~~~Z_{\Gamma_{\pm}^{17,1}+\delta}\to 0,~~~~~(R\to 0).
	\end{align} 
	The models in this class give the 10D non-supersymmetric models in Tables 1 and 2 in the limit $R\to\infty$ and the supersymmetric heterotic models in the limit $R\to 0$. The models in class (2) and class (3) are called interpolating models since they interpolate between two different higher-dimensional string vacua.
	
	\item $\left|\hat{\pi} \right|^2=2~(\text{mod 4}),~~\left(\hat{m},\hat{n} \right) =(1,1)$;\\
	In this class, $\Gamma_{\pm}^{17,1}$ and $\Gamma_{\pm}^{17,1}+\delta$ are written as 
	\begin{align}
	\label{class4Gammapm}
	&\Gamma_{\pm}^{17,1}=
	\left\lbrace p=Z\tilde{\mathcal{E}}\left| \left(\pi,m,n \right)\in\left( \Gamma^{16}_{\pm},2\mathbb{Z},2\mathbb{Z}\right)    \right.  \right\rbrace+
	\left\lbrace p=Z\tilde{\mathcal{E}}\left| \left(\pi,m,n \right)\in\left( \Gamma^{16}_{\pm},2\mathbb{Z}+1,2\mathbb{Z}+1\right)   \right.  \right\rbrace
	\nonumber\\&~~~~~~
	+\left\lbrace p=Z\tilde{\mathcal{E}}\left| \left(\pi,m,n \right)\in\left( \Gamma^{16}_{\mp},2\mathbb{Z},2\mathbb{Z}+1\right)  \right.  \right\rbrace+
	\left\lbrace p=Z\tilde{\mathcal{E}}\left| \left(\pi,m,n \right)\in\left( \Gamma^{16}_{\mp},2\mathbb{Z}+1,2\mathbb{Z}\right)  \right.  \right\rbrace,\\
	\label{class4Gammapmdelta}
	&\Gamma_{\pm}^{17,1}+\delta=
	\left\lbrace p=Z\tilde{\mathcal{E}}\left| \left(\pi,m,n \right)\in\left( \Gamma^{16}_{\pm}+\frac{\hat{\pi}}{2},2\mathbb{Z}+\frac{1}{2},2\mathbb{Z}+\frac{1}{2}\right)\right.  \right\rbrace
	\nonumber\\&~~~~~~~~~~~~~~
	+\left\lbrace p=Z\tilde{\mathcal{E}}\left| \left(\pi,m,n \right)\in\left( \Gamma^{16}_{\pm}+\frac{\hat{\pi}}{2},2\mathbb{Z}-\frac{1}{2},2\mathbb{Z}-\frac{1}{2}\right) \right.  \right\rbrace
	\nonumber\\&~~~~~~~~~~~~~~
	+\left\lbrace p=Z\tilde{\mathcal{E}}\left| \left(\pi,m,n \right)\in\left( \Gamma^{16}_{\mp}+\frac{\hat{\pi}}{2},2\mathbb{Z}+\frac{1}{2},2\mathbb{Z}-\frac{1}{2}\right) \right.  \right\rbrace
	\nonumber\\&~~~~~~~~~~~~~~
	+\left\lbrace p=Z\tilde{\mathcal{E}}\left| \left(\pi,m,n \right)\in\left( \Gamma^{16}_{\mp}+\frac{\hat{\pi}}{2},2\mathbb{Z}-\frac{1}{2},2\mathbb{Z}+\frac{1}{2}\right) \right.  \right\rbrace,
	\end{align}	
	and in the endpoint limits, $Z_{\Gamma_{\pm}^{17,1}}$ and $Z_{\Gamma_{\pm}^{17,1}+\delta}$ behave as follows:
	\begin{align}
	&Z_{\Gamma_{\pm}^{17,1}}\to \frac{R}{\sqrt{\tau_{2}}}\left(\eta\bar{\eta} \right)^{-1}Z_{\Gamma^{16}},
	~~~Z_{\Gamma_{\pm}^{17,1}+\delta}\to 0,~~~~~(R\to \infty),\\
	&Z_{\Gamma_{\pm}^{17,1}}\to \frac{1}{R\sqrt{\tau_{2}}}\left(\eta\bar{\eta} \right)^{-1}Z_{\Gamma^{16}},~~~Z_{\Gamma_{\pm}^{17,1}+\delta}\to 0,~~~~~(R\to 0).
	\end{align} 
	In this class, the supersymmetry is asymptotically restoring in both of the endpoint limits although it is broken at finite values of $R$.
\end{enumerate}

\begin{table}[t]
	\begin{center}
		\begin{tabular}{|c|c|c|c|c|} \hline
			& class (1) & class (2) & class (3) & class (4) \\\hline
			$g_{A}(a)$ & $\pi_{a}\in \Gamma^{16}_{+}(\hat{\pi})$ & $\pi_{a}\in 2\Gamma^{16}$ & $\pi_{a}\in \Gamma^{16}_{+}(\hat{\pi})$ & $\pi_{a}\in 2\Gamma^{16}$  \\
			$g_{\tilde{A}}(\tilde{a})$ & $\pi_{\tilde{a}}\in \Gamma^{16}_{+}(\hat{\pi})$ & $\pi_{\tilde{a}}\in \Gamma^{16}_{+}(\hat{\pi})$ & $\pi_{\tilde{a}}\in 2\Gamma^{16}$ & $\pi_{\tilde{a}}\in 2\Gamma^{16}$  \\
			$g_{D}$ & $g_{D}$ & --- & --- & $g_{D}$  \\\hline
		\end{tabular}
		\caption{The conditions for $g_{A}(a)$, $g_{\tilde{A}}(\tilde{a})$ and $g_{D}$ to be symmetries in the 9D non-supersymmetric heterotic models.}
	\end{center}
\end{table}
Now let us study elements of $D_{(\hat{Z})}(17,1)$ for each of the classes. With $d=1$, $g_{e}(K)$ and $g_{B}(\Theta)$ have no degrees of freedom to make non-trivial transformations. 
So, we focus on $g_{\Gamma^{16}}(W)$, $g_{A}(a)$, $g_{\tilde{A}}(\tilde{a})$ and $g_{D}$. 
Under $g_{\Gamma^{16}}(W)$, $\hat{Z}$ transforms as $\left(\hat{\pi},\hat{m},\hat{n} \right) \to \left(\hat{\pi}W,\hat{m},\hat{n} \right)$, and the choices of $\hat{m}$ and $\hat{n}$ are not changed. Hence, the transitions among the different classes cannot be realized by acting $g_{\Gamma^{16}}(W)$.
For $g_{A}(a)$ the conditions to be in $D_{(\hat{Z})}(17,1)$ are
\begin{align}
\hat{m}\pi_{a}\in 2\Gamma^{16},~~~~\left( \hat{\pi}-\frac{1}{2}\hat{m}\pi_{a}\right) \pi_{a}^{t}=0~(\text{mod 2}).
\end{align}
In classes (1) and (3), in which $\hat{m}=0$, these conditions require that $\pi_{a}$ be in $\Gamma^{16}_{+}(\hat{\pi})$. In classes (2) and (4), in which $\hat{m}=1$, shift parameters must satisfy $\pi_{a}\in 2\Gamma^{16}$ in order for $g_{A}(a)$ to be symmetries. Note that $\pi_{a}\in2\Gamma^{16}$ is a sufficient condition for $\pi_{a}\in\Gamma^{16}_{+}(\hat{\pi})$ and the Wilson line shift symmetry in classes (2) and (4) is more restricted than that in classes (1) and (3). On the other hand, the conditions for $g_{\tilde{A}}(\tilde{a})$ are
\begin{align}
\hat{n}\pi_{a}\in 2\Gamma^{16},~~~~\left( \hat{\pi}-\frac{1}{2}\hat{n}\pi_{a}\right) \pi_{a}^{t}=0~(\text{mod 2}),
\end{align}
and these require $\pi_{a}\in\Gamma^{16}_{+}(\hat{\pi})$ for classes (1) and (2), while $\pi_{a}\in 2\Gamma^{16}$ for classes (2) and (4). As mentioned below \eqref{inversion in hetero}, the inversion $g_{D}$ is a symmetry in the models with the choice $\hat{Z}$ satisfying $\hat{m}=\hat{n}$. With $d=1$, such models belong to class (1) and class (4). The models in class (1) and class (4) being invariant under $g_{D}$ reflects giving the same behaviors in both of the endpoint limits. Table 5 summarizes the (dual) Wilson line shift symmetries and the inversion in each of the four classes.

As mentioned below \eqref{duality transition}, the transitions among the non-supersymmetric models are realized by acting $g\notin D_{(\hat{Z})}(17,1)$ on the models whose T-duality group is $D_{(\hat{Z})}(17,1)$. We study the transitions induced by $g_{A}(a)$, $g_{\tilde{A}}(\tilde{a})$ and $g_{D}$ since these elements allow the transitions from models in a class to those in a different class. To see that, let us define a set $\Gamma^{16}_{+-}(\hat{\pi})$ as
\begin{align}
\Gamma^{16}_{+-}(\hat{\pi})=\left\lbrace \pi\in\Gamma^{16}\left| \pi\cdot \hat{\pi}\in 2\mathbb{Z},~|\pi|^{2}=2~(\text{mod 4})~~\text{or}~~\pi\cdot \hat{\pi}\in 2\mathbb{Z}+1,~|\pi|^{2}=0~(\text{mod 4})\right.  \right\rbrace 
\end{align}
Note that $\hat{\pi}\cdot \pi-|\pi|^{2}/2$ is an odd number for $\pi\in\Gamma^{16}_{+-}(\hat{\pi})$. One can see that the Wilson line shifts $g_{A}(a)$ with $\pi_{a}\in \Gamma_{+}^{16}$ and with $\pi_{a}\in \Gamma_{+-}^{16}$ induce the transition between class (1) and class (3) and that between class (2) and class (4) respectively. On the other hand, the dual Wilson line shifts $g_{\tilde{A}}(\tilde{a})$ with $\pi_{\tilde{a}}\in \Gamma_{+}^{16}$ and with $\pi_{\tilde{a}}\in \Gamma_{+-}^{16}$ induce the transition between class (1) and class (2) and that between class (3) and class (4) respectively. The transition between class (2) and class (3) is realized by the inversion $g_{D}$. Fig. 2 shows the transitions among the different classes induced by $g_{A}(a)$, $g_{\tilde{A}}(\tilde{a})$ and $g_{D}$.

\begin{figure}[t]
	\begin{center}
		\includegraphics[width=15cm]{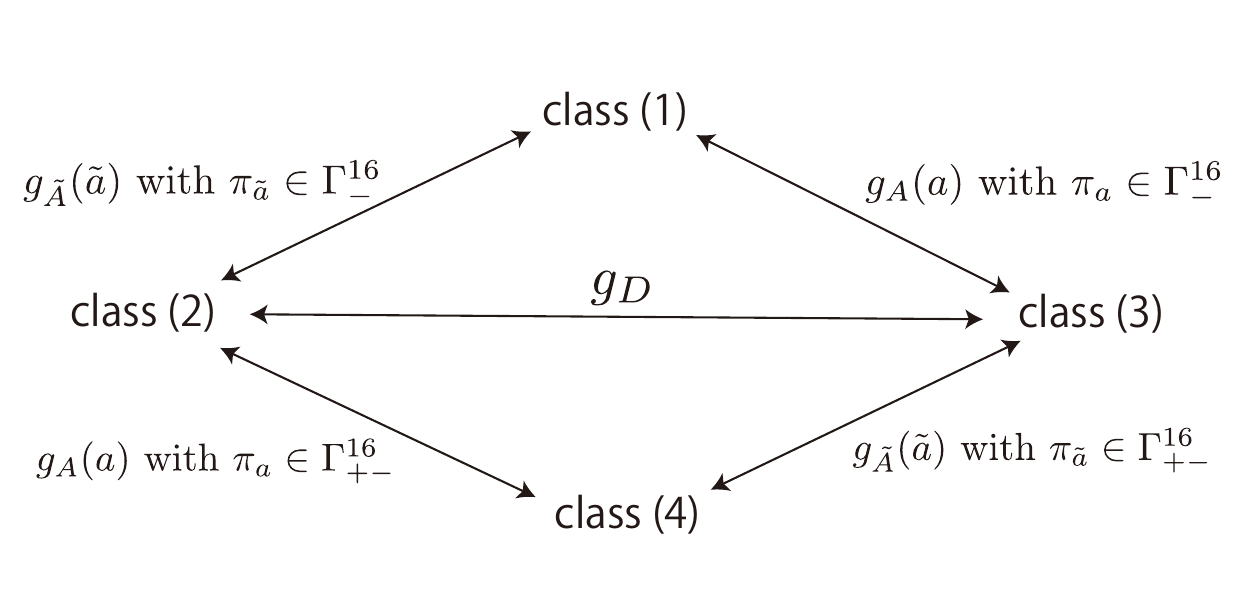}
	\end{center}
	\label{fig:two}
	\vspace{-10mm}
	\caption{An example of the transitions among the non-supersymmetric heterotic models with $d=1$ is shown.}
\end{figure}

\subsubsection{Gauge symmetry enhancement}

At the end of this paper, we briefly discuss special points in the moduli space of the 9D non-supersymmetric heterotic models where the gauge symmetries are enhanced, comparing with the case of the toroidal models. A more detailed analysis is presented for the toroidal heterotic models in \cite{Fraiman:2018ebo,Font:2020rsk} and for the non-supersymmetric models in \cite{Itoyama:2019yst,Itoyama:2020ifw,Itoyama:2021fwc,Itoyama:2021kxp}.
Whatever the value of the moduli takes, there exist massless states satisfying $\ell_{L}= p_{L}=p_{R}=0$, which correspond to a nine-dimensional gravity multiplet and nine-dimensional gauge bosons of $U(1)_{L}^{16}\times U(1)_{l}\times U(1)_{r}$. The additional massless states appear if the moduli satisfy the following conditions which come from $\left|\ell_{L} \right|^{2}+p_{L}^{2}=2$ and $p_{R}=0$:
\begin{align}\label{massless conditions}
n=m\left( R^{2}+\frac{1}{2}|A|^{2}\right)-\pi\cdot A,~~~\left| \pi-mA \right|^{2}+2m^{2}R^{2}=2.
\end{align}
We should note that the conditions \eqref{massless conditions} can be written as
\begin{align}\label{massless conditions dual}
m=n\left( \tilde{R}^{2}+\frac{1}{2}|\tilde{A}|^{2}\right)-\pi\cdot \tilde{A},~~~\left| \pi-n\tilde{A} \right|^{2}+2n^{2}\tilde{R}^{2}=2,
\end{align}
where $\tilde{R}$ and $\tilde{A}$ are defined as
\begin{align}
\tilde{R}=\frac{R}{R^{2}+\frac{1}{2}|A|^{2}},~~~\tilde{A}=-\frac{A}{R^{2}+\frac{1}{2}|A|^{2}}.
\end{align}
In fact, one can check that acting $g_{D}$ on the generalized vierbein \eqref{generalized vierbein hetero} with $d=1$ gives the transformations $R\to \tilde{R}$ and $A\to \tilde{A}$ accompanied with an appropriate $O(17,\mathbb{R})\times O(1,\mathbb{R})$ rotation.

First, we study the conditions under which $U(1)_{L}^{16}$ is enhanced to $SO(32)$ or $E_{8}\times E_{8}$. For a while, we assume that $R$ takes a generic value. The enhancement is then realized either by the states with $m=0$ or with $n=0$, for which the conditions \eqref{massless conditions} and \eqref{massless conditions dual} are respectively
\begin{subequations}
	\begin{align}
	\label{conditions with m=0}
	n&=-\pi\cdot A,~~~\left| \pi\right|^{2}=2,~~~\text{with $m=0$},\\
	\label{conditions with n=0}
	m&=-\pi\cdot \tilde{A},~~~\left| \pi\right|^{2}=2,~~~\text{with $n=0$}.
	\end{align}
\end{subequations}
For the toroidal models, these conditions imply that the gauge symmetry is enhanced to $SO(32)$ or $E_{8}\times E_{8}$ when $A\in \Gamma_{g}^{*}$ or $\tilde{A}\in \Gamma_{g}^{*}$. Here $\Gamma_{g}^{*}$ denotes the weight lattice of $SO(32)$ or $E_{8}\times  E_{8}$. For the non-supersymmetric models, the situation is different since $m$ and $n$ do not necessarily take any integer numbers and $\Gamma^{16}$ is split into $\Gamma^{16}_{+}$ and $\Gamma^{16}_{-}$ by a shift-vector, depending on the spacetime $SO(8)$ representations. Noting that spacetime vectors live in $\Gamma^{17,1}_{+}$, we find the following conditions for each of the four classes;
\begin{itemize}
	\item class (1);\\
	The structure of $\Gamma_{+}^{17,1}$ is given by \eqref{class1Gammapm}, and there is no point with the enhancement to $SO(32)$ or $E_{8}\times E_{8}$ unless all $\pi\in\Gamma^{16}$ that satisfy $|\pi|^{2}=2$ are included in $\Gamma^{16}_{+}$. Such $\Gamma^{16}_{+}$ is obtained in the $Spin(32)/\mathbb{Z}_{2}$ models by choosing $\hat{\pi}=2e_{i}$ with $e_{i}=(0,\cdots,1,\cdots,0)$ being the $i$-th orthonormal vector.
	
	\item class (2);\\
	Noting that $n\in2\mathbb{Z}$ for $\pi\in\Gamma_{+}^{16}$ and $n\in2\mathbb{Z}+1$ for $\pi\in\Gamma_{-}^{16}$ in $\Gamma_{+}^{17,1}$, we find that the conditions \eqref{conditions with m=0} are fulfilled with $A=\hat{\pi}$. Thus, the enhancement to $E_{8}\times E_{8}$ or $SO(32)$ is reached by the states with $m=0$. 
	
	\item class (3);\\
	In class (3), the roles of $m$ and $n$ are switched compared to in class (2). The enhancement is hence realized by the conditions \eqref{conditions with n=0} with $\tilde{A}=\hat{\pi}$, and the states with $n=0$ can become the nonzero roots of the non-Abelian gauge group.
	
	\item class (4);\\
	From \eqref{class4Gammapm}, it turns out that the enhancement to $E_{8}\times E_{8}$ or $SO(32)$ is reached by both of the conditions \eqref{conditions with m=0} with $A=\hat{\pi}$ and \eqref{conditions with n=0} with $\tilde{A}=\hat{\pi}$.
\end{itemize}
In class (2) and class (4), in order for the enhancement to be maintained, the Wilson lines must be shifted by $\pi_{a}\in2\Gamma^{16}$, not by $\pi_{a}\in\Gamma^{16}$. We can apply the same argument for the dual Wilson lines to the cases in class (3) and class (4). These results are consistent with that the Wilson line shift symmetries in class (2) and (4) and the dual Wilson line shift symmetries in class (3) and (4) require $\pi_{a}\in2\Gamma^{16}$ and $\pi_{\tilde{a}}\in2\Gamma^{16}$ respectively, as shown in Table 5.

Next, let us study the possibility of the enhancement $U(1)_{l}\to SU(2)$ under an assumption that the Wilson line takes a generic value, i.e., $\pi\cdot A$ is a generic real value unless $\pi=0$. Hence, the states that can be the nonzero roots of $SU(2)$ are with $\pi=0$, and the conditions \eqref{massless conditions} for such states are
\begin{align}
n=m\left( R^{2}+\frac{1}{2}|A|^{2}\right),~~~2mn=2.
\end{align}
These conditions lead to $m=n=\pm1$ and $R^{2}+\frac{1}{2}|A|^{2}=1$. The latter implies $R=\tilde{R}$ and $A=-\tilde{A}$, that is, the fixed points of the inversion $g_{D}$. Focusing on the structures of $\Gamma_{+}^{17,1}$ in each class, we find that the spacetime vectors with $\pi=0$ and $m=n=\pm1$ exist only in class (1) and class (4). Namely, the enhancement $U(1)_{l}\to SU(2)$ can be realized at the $g_{D}$-fixed points in the moduli space of the models in class (1) and class (4). This result reflects that only the models in class (1) and class (4) are invariant under $g_{D}$, as shown in Table 5. 

\section*{Acknowledgments}
The work of HI is supported in part by JSPS KAKENHI Grant Number 19K03828 and by the Osaka City University (OCU) Strategic Research Grant 2020 for priority area (OCU-SRG2019 TPR01). The work of SN is supported in part by JSPS KAKENHI Grant Number 21J15497. The work of YK is supported by the establishment of university fellowships towards the creation of science technology innovation.


\appendix

\section{Lattices and characters}\label{appendixA}

Irreducible representations of $SO(2n)$ can be classified into the four conjugacy classes:
\begin{itemize}
	\item The trivial conjugacy class (the root lattice):
	\begin{align}
	\Gamma^{(n)}_{g}=\left\lbrace \left( n_1,\cdots,n_{n}\right) \left|~ n_{i}\in \mathbb{Z},~\sum_{i=1}^{n}n_{i}\in 2\mathbb{Z} \right. \right\rbrace .
	\end{align}
	\item The vector conjugacy class:
	\begin{align}
	\Gamma^{(n)}_{v}=\left\lbrace \left( n_1,\cdots,n_{n}\right) \left|~ n_{i}\in \mathbb{Z},~\sum_{i=1}^{n}n_{i}\in 2\mathbb{Z} +1\right. \right\rbrace .
	\end{align}
	\item The spinor conjugacy class:
	\begin{align}
	\Gamma^{(n)}_{s}=\left\lbrace \left( n_1+\frac{1}{2},\cdots,n_{n}+\frac{1}{2}\right) \left|~ n_{i}\in \mathbb{Z},~\sum_{i=1}^{n}n_{i}\in 2\mathbb{Z} \right. \right\rbrace .
	\end{align}
	\item The conjugate spinor conjugacy class:
	\begin{align}
	\Gamma^{(n)}_{c}=\left\lbrace \left( n_1+\frac{1}{2},\cdots,n_{n}+\frac{1}{2}\right) \left|~ n_{i}\in \mathbb{Z},~\sum_{i=1}^{n}n_{i}\in 2\mathbb{Z}+1 \right. \right\rbrace .
	\end{align}
\end{itemize}
The weight lattice of $SO(2n)$, which is dual to $\Gamma^{(n)}_{g}$, is given by the sum of the four conjugacy classes:
\begin{align}
\Gamma^{(n)}_{w}=\Gamma^{(n)}_{g}+\Gamma^{(n)}_{v}+\Gamma^{(n)}_{s}+\Gamma^{(n)}_{c}.
\end{align}

Modular invariance of the partition functions of the 10D supersymmetric heterotic string models requires that the internal momenta should live in an even self-dual Euclidean lattice. In 16-dimensions, only two such lattices exist. One of them is the root lattice of $E_{8}\times E_{8}$,
\begin{align}
\Gamma^{16}=\left( \Gamma^{(8)}_{g}+\Gamma^{(8)}_{s}\right) \times \left( \Gamma^{(8)}_{g}+\Gamma^{(8)}_{s}\right),
\end{align}
and the other is that of $Spin(32)/\mathbb{Z}_{2}$ which is expressed as the sum of the trivial and spinor conjugacy classes of $SO(32)$:
\begin{align}
\Gamma^{16}= \Gamma^{(16)}_{g}+\Gamma^{(16)}_{s}.
\end{align}

The $SO(2n)$ characters of the corresponding conjugacy classes are defined as
\begin{align}
O_{2n}
&=\frac{1}{\eta^{n}} \sum_{\pi\in \Gamma^{(n)}_{g}}q^{\frac{1}{2}|\pi|^2}
=\frac{1}{2\eta^{n}}\left( \vartheta^{n}
\begin{bmatrix} 
0\\ 
0\\ 
\end{bmatrix}(0,\tau)+ \vartheta^{n}
\begin{bmatrix} 
0\\ 
1/2\\ 
\end{bmatrix}(0,\tau)
\right),\\
V_{2n}
&=\frac{1}{\eta^{n}} \sum_{\pi\in \Gamma^{(n)}_{v}}q^{\frac{1}{2}|\pi|^2}
=\frac{1}{2\eta^{n}}\left( \vartheta^{n}
\begin{bmatrix} 
0\\ 
0\\ 
\end{bmatrix}(0,\tau)- \vartheta^{n}
\begin{bmatrix} 
0\\ 
1/2\\ 
\end{bmatrix}(0,\tau)
\right),\\
S_{2n}
&=\frac{1}{\eta^{n}} \sum_{\pi\in \Gamma^{(n)}_{s}}q^{\frac{1}{2}|\pi|^2}
=\frac{1}{2\eta^{n}}\left( \vartheta^{n}
\begin{bmatrix} 
1/2\\ 
0\\ 
\end{bmatrix}(0,\tau)+ \vartheta^{n}
\begin{bmatrix} 
1/2\\ 
1/2\\ 
\end{bmatrix}(0,\tau)
\right),\\
C_{2n}
&=\frac{1}{\eta^{n}} \sum_{\pi\in \Gamma^{(n)}_{c}}q^{\frac{1}{2}|\pi|^2}
=\frac{1}{2\eta^{n}}\left( \vartheta^{n}
\begin{bmatrix} 
1/2\\ 
0\\ 
\end{bmatrix}(0,\tau)-\vartheta^{n}
\begin{bmatrix} 
1/2\\ 
1/2\\ 
\end{bmatrix}(0,\tau)
\right),
\end{align}
where the Dedekind eta function and the theta function with characteristics are defined as
\begin{align}
\eta(\tau)&=q^{1/24}\prod_{n=1}^{\infty}\left( 1-q^{n}\right),\\
\vartheta
\begin{bmatrix} 
\alpha\\ 
\beta\\ 
\end{bmatrix}(z,\tau)&=\sum_{n=-\infty}^{\infty}\exp\left( \pi i (n+\alpha)^2 \tau +2\pi i (n+\alpha)(z+\beta) \right). 
\end{align} 

\section{Invariance under $\tau\to\tau+1$}\label{check modular inv}

In this appendix, we check the invariance of the partition functions \eqref{non-susy typeiiB} and \eqref{non-susy hetero} under $\tau \to \tau+1$. In particular, we show that the combinations of the $SO(8)$ characters and $\Gamma^{d_{L},d_{R}}_{\pm}+\delta$ in the twisted sectors must be taken appropriately. To do so, we should note that the Dedekind eta function and the $SO(8)$ characters transform under $\tau \to \tau+1$ as follows:
\begin{align}
\eta(\tau)&\to \eta(\tau+1)=e^{\frac{\pi i}{12}}\eta(\tau),\\
\left( O_{8},V_{8},S_{8},C_{8}\right)&\to e^{-\frac{\pi i}{3}}\left( O_{8},-V_{8},-S_{8},-C_{8}\right).
\end{align}
The products of the $SO(8)$ characters and $\eta^{8}$ have an eigenvalue $-1$ only for the trivial conjugacy class:
\begin{align}
\eta^{-8}\left( O_{8},V_{8},S_{8},C_{8}\right)\xrightarrow{\tau\to\tau+1} \eta^{-8}\left( -O_{8},V_{8},S_{8},C_{8}\right).
\end{align}
Then, the untwisted sectors, which include neither $O_{8}$ nor $\bar{O}_{8}$, are obviously invariant since $p$ is in the even lattice. In the twisted sectors, the momenta are shifted by $\delta$ and we get the following phase from $Z_{\Gamma_{\pm}^{d_L,d_{R}}+\delta}$ under $\tau\to\tau+1$, excluding the phase which comes from $\eta^{-d_{L}}\bar{\eta}^{-d_{R}}$:
\begin{align}\label{phase from lattice}
e^{\pi i \left( p^{2}+\delta^{2}+2p\cdot\delta\right)}.
\end{align}
If $\delta^{2}$ is even and $p\in\Gamma_{+}^{d_{L}d_{R}}$, or $\delta^{2}$ is odd and $p\in\Gamma_{-}^{d_{L}d_{R}}$ then the phase \eqref{phase from lattice} is 1. Thus, $Z_{\Gamma_{+}^{d_{L}d_{R}}+\delta}$ with $\delta^{2}$ even or $Z_{\Gamma_{-}^{d_{L}d_{R}}+\delta}$ with $\delta^{2}$ odd must be accompanied with $\left(  O_{8}\bar{O}_{8}+C_{8}\bar{C}_{8}\right)$ in the type IIB case and with $\bar{C}_{8}$ in the heterotic case. On the other hand, the phase \eqref{phase from lattice} is $-1$ if $\delta^{2}$ is even and $p\in\Gamma_{-}^{d_{L}d_{R}}$, or $\delta^{2}$ is odd and $p\in\Gamma_{+}^{d_{L}d_{R}}$. Thus, $Z_{\Gamma_{-}^{d_{L}d_{R}}+\delta}$ with $\delta^{2}$ even or $Z_{\Gamma_{+}^{d_{L}d_{R}}+\delta}$ with $\delta^{2}$ odd must be accompanied with $\left(  O_{8}\bar{C}_{8}+C_{8}\bar{O}_{8}\right)$ in the type IIB case and with $\bar{O}_{8}$ in the heterotic case. As a result, we see that the partition functions \eqref{non-susy typeiiB} and \eqref{non-susy hetero} have the appropriate combinations of the $SO(8)$ characters and $\Gamma^{d_{L},d_{R}}_{\pm}+\delta$, and hence they are invariant under the shift $\tau\to\tau+1$.


\end{document}